\documentclass[12pt]{article}

\usepackage{scicite}
\usepackage{times}

\newenvironment{sciabstract}{%
\begin{quote} \bf}
{\end{quote}}

\usepackage{amsmath}
\usepackage{amsfonts}
\usepackage{amssymb}
\usepackage{graphicx}
\usepackage{import}
\usepackage{dcolumn}
\usepackage{bm,bbm}
\usepackage{float}
\usepackage{placeins}
\usepackage{subfiles}
\usepackage{xcolor}
\usepackage{hyperref}
\makeatletter
\newcommand\setcurrentname[1]{\def\@currentlabelname{#1}}
\makeatother
\usepackage[capitalise]{cleveref}
\usepackage[toc,page]{appendix}
\usepackage{array,makecell}

\usepackage{hyperref}
\usepackage{algpseudocode}
\graphicspath{{figures/}}
\usepackage{caption}
\creflabelformat{equation}{#2(#1)#3}

\usepackage[T1]{fontenc}
\fontfamily{cmr}\selectfont
\usepackage{titling}
\usepackage{authblk}

\usepackage{titlesec}
\usepackage[letterpaper, total={7.25in, 9.5in}]{geometry}

\usepackage{verbatim}

\title{Irreversibility in Bacterial Regulatory Networks} 

\author
{Yi Zhao,${}^{1,2\dagger}$ Thomas P. Wytock,${}^{1,2\dagger}$, Kimberly A. Reynolds,${}^{3,4}$ Adilson E. Motter${}^{1,2,5,6\ast}$\\
\normalsize{${}^{1}$Department of Physics and Astronomy, Northwestern University,}\\ 
\normalsize{Evanston, IL 60208, USA.}\\
\normalsize{${}^{2}$Center for Network Dynamics, Northwestern University,}\\
\normalsize{Evanston, IL 60208, USA.}\\
\normalsize{${}^{3}$The Green Center for Systems Biology -- Lyda Hill Department of Bioinformatics,}\\ 
\normalsize{University of Texas Southwestern Medical Center, Dallas, TX 75390, USA.}\\
\normalsize{${}^{4}$Department of Biophysics, University of Texas Southwestern Medical Center,}\\ \normalsize{Dallas, TX 75390, USA.}\\
\normalsize{${}^{5}$Northwestern Institute on Complex Systems, Northwestern University,}\\
\normalsize{Evanston, IL 60208, USA.}\\
\normalsize{${}^{6}$National Institute for Theory and Mathematics in Biology,}\\ 
\normalsize{Evanston, IL 60208, USA.}
\\
\normalsize{$^\dagger$These authors contributed equally to this work.}\\
\normalsize{$^\ast$To whom correspondence should be addressed; E-mail:  motter@northwestern.edu.}
\vspace{-0.4cm}}

\date{}

\begin{document}

\baselineskip24pt

\maketitle 
\vspace{-1cm}
\noindent {\bf Short title:}
Irreversibility in Bacterial Regulatory Networks

\noindent {\bf Sentence summary:}

\noindent  Transient gene perturbations cause heritable transcriptomic changes in \textit{E. coli}; thus one genotype can have multiple phenotypes.

\begin{sciabstract}
Irreversibility, in which a transient perturbation leaves a system in a new state, is an emergent property in systems of interacting entities.
This property has well-established implications in statistical physics but remains underexplored in biological networks, especially for bacteria and other prokaryotes whose regulation of gene expression occurs predominantly at the transcriptional level. 
Focusing on the reconstructed regulatory network of \emph{Escherichia coli}, we
examine network responses to transient single-gene perturbations.
We predict irreversibility in numerous cases and find that the incidence of irreversibility increases with the proximity of the perturbed gene to positive circuits in the network.
Comparison with experimental data suggests a connection between the predicted irreversibility to transient perturbations and the evolutionary response to permanent perturbations.
\end{sciabstract}

\clearpage
\newpage

\section*{Introduction}
A common goal in both statistical physics and systems biology is to
connect the attributes of microscopic entities with 
observable macroscopic properties. 
Of particular interest are macroscopic properties that are emergent---including pattern formation~\cite{Koch1994} and synchronization~\cite{Winfree2001}---because 
they arise from interactions between system entities and can therefore enable new system-level functionality.
In statistical physics, an important emergent property is the irreversibility
of macroscopic processes~\cite{Lynn2022}, where entropy---a state function---can increase irreversibly despite 
the time-reversibility of the microscopic dynamics.
A related property is hysteresis \cite{Keim2019}, where a cyclic (reversible) change of a variable leads to a persistent change in the state of the system.

In molecular biophysics, a central dogma \cite{Crick1970} posits that phenotype is determined by genotype and would thus be reversible.
That is, identical DNA sequences would yield identical observable characteristics, which are assumed to arise from the proteins of the cell. The dogma allows for the possibility that multiple DNA sequences can map to the same protein amino acid sequence through synonymous codon usage, but multiple amino acids are not assigned to the same codon. Thus, rigorous adherence to this dogma cannot account for eukaryotic processes like organismal development \cite{Villarreal2012}, cell differentiation \cite{Huang2005}, cell reprogramming \cite{Zhou2011,Zanudo2015,Zhu2022}, and nongenetic aspects of aging \cite{Lopez-Otin2013,He2017}, which may nevertheless be attributed to epigenetics (e.g., histone modifications and DNA methylation).
It also does not account for environmentally induced switches in prokaryotic systems, such as stalking in \textit{Caulobacter crescentus}~\cite{Shapiro1971}, sporulation in \textit{Bacillus subtilus}~\cite{Veening2008}, and the \textit{lac} and \textit{mar} operons in \textit{Escherichia coli}~\cite{Ozbudak2004,Prajapat2015}---despite epigenetic mechanisms being largely absent in prokaryotes.
The extent to which the phenotype-genotype correspondence holds for prokaryotes beyond specific cases remains an outstanding question.
It is thus natural to ask how and when phenotype, which is to first approximation a state function of gene activity,
can change irreversibly in prokaryotes even in response to transient microscopic (e.g., single-gene)  perturbations. 
Here, we will consider a change to be irreversible when such a short-lived perturbation leads to a long-lasting heritable phenotypic change that persists across multiple cell divisions.
In a deterministic mathematical model, these changes would be associated with transitions between stable states and hence be permanent.
 While dissipation is necessary for the existence of stable states, it is not sufficient for irreversible changes to occur, 
since the final stable state may be the same as the original one.
By considering genetic rather than environmental perturbations,
we can address
whether transcriptional regulation itself deviates from the central dogma and does so even in the absence of environmental inducers.

\begin{figure}[t]\centering
\includegraphics[width=\textwidth]{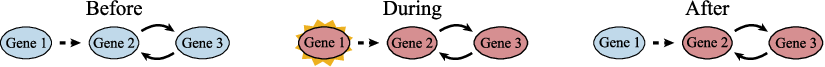}
\caption{\textbf{Example of the mechanism for irreversibility in a simple network of activating relationships.} Before perturbation, genes are in the ``OFF'' state (blue color). During the perturbation, gene 1 is perturbed (yellow star background), which turns ``ON'' genes 2 and 3 (red). After the perturbation, gene 1 is restored to its initial OFF state, but genes 2 and 3 remain ON.
}
\label{fig:nec-suff}
\end{figure}

In this paper, we study the prevalence and network mechanisms of irreversibility in the gene regulatory network of \emph{E. coli}, which is the model organism with the most complete reconstructed network of this kind currently available. 
Because this reconstruction is the union of potential regulatory interactions whose presence may depend on the environmental conditions, we consider irreversibility across a range of representative sets of interactions.
Using Boolean dynamics modeling, we examine the impact of transient single-gene perturbations on the activity of the other genes.
We predict that 
transient knockouts (KOs) and transient overexpressions (OEs) of individual genes
in the central part of the regulatory network
commonly  
result in irreversible changes in the states of other genes in the network. 
Our results identify positive circuits (network cycles with an even number of repressive interactions) as the relevant network structures underlying the multistability necessary for irreversibility.
Mechanistically,
a transient perturbation may lead to permanent changes when
it alters
(directly or indirectly) the state of one or more multistable positive circuits (\cref{fig:nec-suff}).
This condition is more frequently satisfied the closer the perturbed gene is to a downstream positive circuit, and thus the likelihood of a gene being irreversible increases with its proximity to (or membership in) a nontrivial strongly connected component (SCC) of the network. 
Comparing with existing data on adaptive evolution experiments, our predictions
support the hypothesis that the genes remaining in different states following adaptive evolution to a (permanent) gene KO are largely determined by those that respond irreversibly to a transient KO of the same gene.
The results point to specific candidates for observing irreversibility and reveal its the prevalence in prokaryotes despite 
the tight transcription-to-translation
coupling and
the still largely unresolved role of epigenetic mechanisms in such organisms.

\section*{Results}

\paragraph*{Network modeling approach.}
We focus our analysis on the transcriptional regulatory network from the RegulonDB database, 
$G(V, E)$, which is reconstructed from empirical data and 
includes $|V|=1{,}859$ genes (nodes) and $|E|=5{,}119$ pairwise signed interactions (edges)~\cite{Santos-Zavaleta2018}. 
We retain only activating (positive sign) and repressing (negative sign) interactions, leaving out 148 edges with dual or unknown function.
Recognizing that the subset of regulatory interactions present (but not their polarity) may vary depending on the cultivation conditions, we examine the average irreversibility propensity across a representative ensemble of dynamical rules formed by these interactions. 
In order to ensure that our analysis is conducted on a connected component of the network, we identified all origons.
An origon is a subnetwork that starts at a root node with no incoming edges (other than autoregulatory ones) and includes all downstream nodes and edges that can be reached from the root node by following directed paths~\cite{Balazsi2005}.

We analyze the largest origon, which is rooted at gene \emph{phoB} and consists of a 1,406-node subnetwork;  the results would be similar for all other large origons since they are accounted for by related \textit{core networks} (\cref{tab:origon_overlap}). 
The \textit{phoB} origon is also the largest subnetwork reachable from any individual node in the RegulonDB model.
The origon core is the subnetwork $G'(V',E')$ that remains after recursively removing the nodes with zero outgoing edges.
According to this procedure, which generalizes the metabolic network trimming used in Ref.~\cite{Samal2008}, the core network shares similarities with the concept of $k$-core applied to the outgoing edges.
Our notion of core network is nevertheless different from other generalizations of $k$-core on directed networks~\cite{Azimi-Tafreshi2013} as the procedure here is tailored to capture the irreversibility of the entire origon.
This is the case because nodes outside the core have no impact on the state of upstream nodes (including all nodes in the core) and have reversible impacts on the state of downstream nodes.
In the case of the \textit{phoB} origon, the core network consists 
of $|V'|=87$ nodes and $|E'|= 290$ edges. 
Given that the number of Boolean network states scales as $2^{|V'|}$, 
our focus on the core also leads to a dimension reduction
that helps circumvent a combinatorial explosion in
simulations of the dynamics.

The network dynamics are modeled using a Boolean framework \cite{Kauffman1993,Wang2012} on the core network
with nodes $u = 1,\ldots,|V'|$ and edges $E' \subseteq V' \times V'$.
Edges are denoted by (directed) ordered pairs $(v, u)$ indicating 
that the gene associated with tail node $v$ regulates the gene 
associated with head node $u$. 
The \textit{polarity} of $(v, u)$, $W(v, u)$, is $+1$ or $-1$ indicating activation or repression.
The state of the network at time $t$ is indicated by $ \mathbf{x}^{t}=(x_u^{t})$, where $x_u^{t} \in \{0,1\}$ is the Boolean state of gene $u$.
Each $x_u^{t}$ is assumed to evolve according to a deterministic Boolean function $B_u : \{0,1\}^{V'} \rightarrow \{0,1\}$,
\begin{equation}
 x^{t+1}_u =B_u(\mathbf{x}^{t}), \label{eq:Bool}
 \vspace{-0.1cm}
\end{equation}
which accounts for the $k^+_u$ edges incident on $u$ in $G'$ and their polarities by obeying three network consistency constraints. 
The constraints are edge consistency 
(nodes with states on the right-hand side in \cref{eq:Bool} must connect to the node $u$), 
edge essentiality 
(all nodes on the right-hand side are necessary to determine $x^{t+1}_u$), 
and sign consistency 
($x^{t}_v$ or $\bar{x}^{t}_v$ appears on the right-hand side if $v$ activates or inhibits $u$).
Here we use that the negation of $x_v=0$ is $\bar{x}_v=1$ and the negation of $x_v=1$ is $\bar{x}_v=0$.
As a consequence, $B_u(\mathbf{x})$ can be written as a sum of products of $x_v$ and/or $\bar{x}_v$ (modulus 1) for all $v$ with edges incident on $u$, where $x_v$ ($\bar{x}_v$) appears and does so once if the polarity $W(v, u)$ is positive (negative)~\cite{he2016stratification}. 
For additional details and the special case of autorepression, we refer to the \nameref{sec:methods}.

\begin{figure}[t!]\centering
\includegraphics[width=0.75\textwidth]{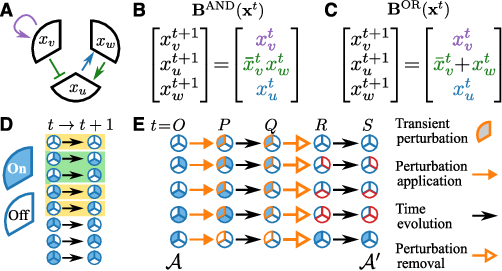}
\caption{\textbf{Finding irreversible perturbations in a Boolean gene regulatory network of 3 nodes.}
(\textbf{A}) Representation of a three-gene network, where the sectors correspond to genes $u$, $v$, and $w$ with states $x_u$, $x_v$, and $x_w$. Here, as in all subsequent network figures, pointed arrowheads indicate activating relationships and flat arrowheads indicate repressive relationships.
(\textbf{B} and \textbf{C}) Boolean functions consistent with the network edges and polarities in (A), as indicated by text colors and negations (bars), respectively. 
The functions $\mathbf{B}^{\rm AND}$ and $\mathbf{B}^{\rm OR}$---labeled according to the  function assigned to update node $u$---provide rules for synchronous updates of the node states at each time $t$.
(\textbf{D}) State transitions for the update rules $\mathbf{B}^{\rm AND}$ where sector colors indicate the node state. The yellow and green backgrounds indicate fixed-point and period-2 attractors, respectively.
(\textbf{E}) Application and removal of a perturbation to each attractor state. This transient perturbation can leave the network in a different attractor,  with the altered node states between the initial attractor $\mathcal{A}$ and final attractor $\mathcal{A}'$ indicated by a red outline.
}
\label{fig:illu}
\end{figure}

As an example of consistent update rules, consider the regulatory relationships illustrated in the 3-gene network of \cref{fig:illu}A.
Because node $u$ has two incoming edges ($k_u^+=2$), there are 2
feasible functions that satisfy the network consistency constraints, $\mathbf{B}^{\rm AND} = (B_u^{\rm AND})$ and $\mathbf{B}^{\rm OR}= (B_u^{\rm OR})$  (\cref{fig:illu}B and \cref{fig:illu}C, respectively).
Both rules have multiple attractors $\mathcal{A}$, which are fixed-point or periodic orbits $\{\mathbf{x}_{\mathcal{A}}^t\} \big|_{t=1}^{T_{\!\mathcal{A}}}$, $T_{\!\mathcal{A}}\geq 1$, to which the other states converge over time.
\Cref{fig:illu}D shows the state transitions and attractors associated with $\mathbf{B}^{\rm AND}$. 
The attractors form the starting point for identifying irreversible transient perturbations as illustrated in \cref{fig:illu}E.
Starting from each attractor ($t=O$), 
each $x_u$ is perturbed independently from $1$ to $0$ for KOs and from $0$ to $1$ for OEs ($t=P$).
The states are allowed to evolve under the perturbation until reaching a new attractor ($t=Q$). 
The perturbation is 
removed upon reaching the attractor ($t=R$), and the states evolve to the final attractor $\mathcal{A}'$ ($t=S$).  
If $\mathcal{A}' \neq \mathcal{A}$,
we classify the transient KO or OE 
as an \textit{irreversible perturbation} 
and refer to genes that differ between attractors as \textit{irreversible response genes}.

\begin{figure}
    \centering
    \includegraphics[width=\textwidth]{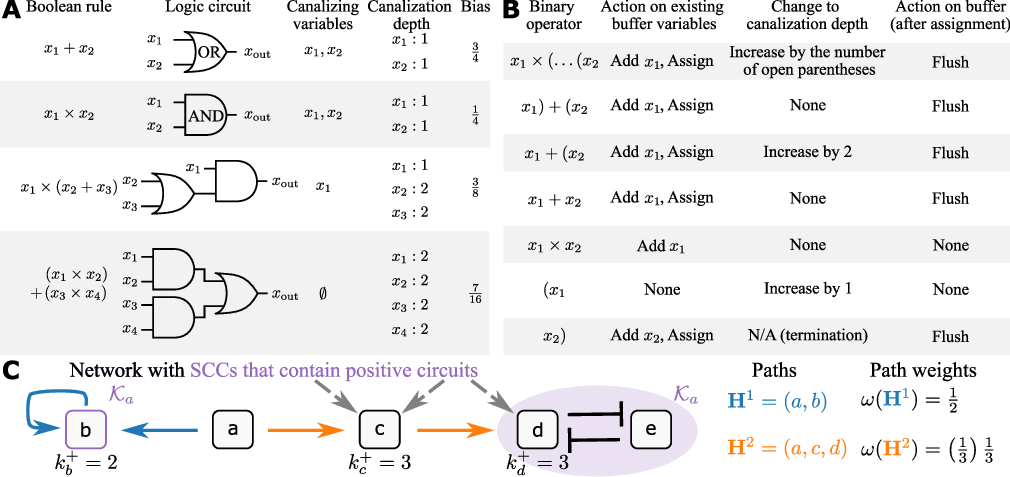}
    \caption{ \textbf{Example calculations of the key Boolean concepts underlying the network ensemble.} 
    (\textbf{A}) Examples of Boolean rules, their logic circuit representation, canalizing variables, variable canalization depth, and rule bias.
    (\textbf{B}) Scheme for calculating the canalization depth of the variables. N/A, Not Applicable. Each row of the table explains how the binary operator determines the assignment of the canalization depth of the variables.
    (\textbf{C}) Illustration of the path weight calculation for network paths to SCCs that contain positive circuits. 
    }
    \label{fig:bool}
\end{figure}

To proceed with the analysis of irreversibility in our model, 
we must first specify the rules of the core regulatory network.
Because the number of possible rules is too large to simulate exhaustively (see \nameref{sec:methods}),
we developed an algorithm to \textit{sample} the ensemble of rules based on key qualities of empirical and Boolean regulatory networks. 
In \cref{fig:bool}A, we illustrate key properties of Boolean networks relevant to our simulations.
Regulatory networks have been empirically observed to have a nested canalizing structure~\cite{Harris2002,Kauffman2003,Moreira2005}, which occurs when  the state of one incident node determines the output regardless of the state of the remaining incident nodes. Mathematically, this condition is expressed as
$x_v = x^*_v$ implies $B_u(\mathbf{x}|_{x_v=x^*_v}) = x^*_u$
independently of the states of a set of one or more other nodes incident on $u$.
As seen in the first two rows of \cref{fig:bool}A, both variables in the simple AND ($\times$) and OR ($+$) are canalizing, while higher order Boolean functions  with higher levels of nesting are shown in the third and fourth rows.
Here, we quantify the nestedness by calculating the expected number of variable states needed to determine the output of $B_u$, which we refer to as the \textit{canalization depth}. 
This is calculated by expressing the Boolean rules in simplest form using the Quine-McCluskey algorithm~\cite{Quine1955,McCluskey1956}.
Briefly, we break down the simplified rule into binary operators and  read each operator from left to right, while keeping track of the canalization depth and a list of variables (i.e., buffer) whose depth is to be assigned.
\Cref{fig:bool}B provides instructions on how to update the canalization depth and list of variables when each operator is encountered. 
For each pair of inputs, we decide whether or not (i) to assign the depth to each variable in the buffer, (ii) to increase the canalization depth, and/or (iii) to empty the buffer.
We also account for the \textit{rule bias}, which is the probability of updating to $1$, as this quantity has been shown to play a role in determining the response to perturbations in random Boolean networks~\cite{Pomerance2009,Squires2014}.

It is therefore natural to sample the ensemble using a \textit{nestedness parameter} $r$ and a \textit{bias parameter} $s$.  Each Boolean function has a list of Boolean variables determined by the network consistency constraints leaving the $k^+_u - 1$ binary operations for us to specify. We assign the nesting operator ``$\times (\,$'' between the variables with probability $r$. In the remaining $1-r$ probability, we assign the $+$ operator with probability $s$ and the $\times$  operator with probability $1-s$. Thus, the binary operators $\times$, $+$, and $\times (\,$ appear with probabilities $(1-r)(1-s)$, $(1-r)s$, and $r$, respectively. 
 With these parameters defined, we can enumerate all rules of $k^+_u$ variables, assign a probability of randomly obtaining each rule, and determine the canalization depth of each variable in each rule (according to the procedure in \cref{fig:bool}B).
By first averaging the canalization depth over the variables in each rule and subsequently weighting by the probability of obtaining each rule, we obtain an expression for the \textit{average canalization depth}.
The parameters $r$ and $s$ also can be used to relate our representation of the rules---which focuses on a core network of densely connected cycles and explicitly accounts for the polarity of each edge in the network---to representations that treat rule inputs as statistically independent~\cite{Tripathi2020,Tripathi2023} and/or are agnostic of edge polarities~\cite{Pomerance2009,Squires2014}.

\begin{figure*}[t]
    \centering
    \includegraphics[width=\textwidth]{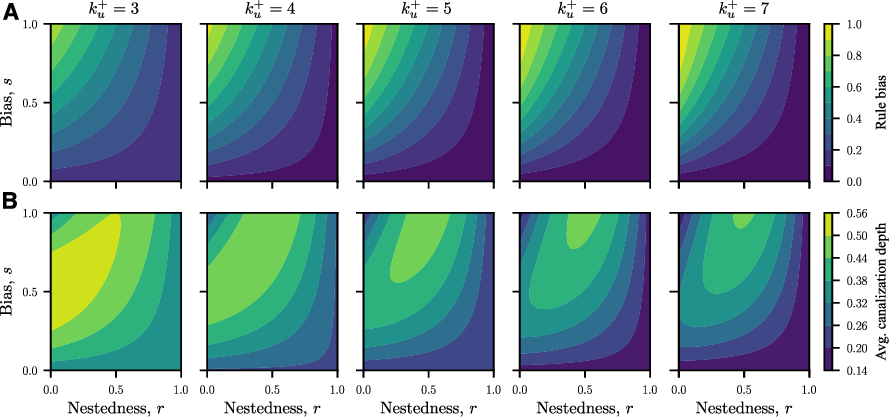}
    \caption{\textbf{Parameter dependence of the update rules.} (\textbf{A}) Rule bias and (\textbf{B}) average canalization depth 
    for $k^+_u=3,...,7$. 
    The rule properties are expressed as functions of the nestedness parameter $r$, which is the probability of joining two inputs with the $ \times ( $ operator, and bias parameter $s$, which is the probability of joining two inputs with the $+$ operator.
    }
    \label{fig:rule-contours}
\end{figure*}

\Cref{fig:rule-contours}A and \cref{fig:rule-contours}B show, respectively, the rule bias and average canalization depth as functions of $r$ and $s$.
These quantities are determined by 
enumerating the $2^{k^+_u-1}$ possible input combinations and their associated probabilities.
In the cases ($r\!=\!0$, $s\!=\!1$), ($s\!=\!0$ $\forall r$),  and ($r\!=\!1$ $\forall s$), the rules are fully canalizing, since the first joins all pairs with a $+$ operator and the latter two join all pairs with a $\times$ operator. 
The rule bias and canalizing state for each input are, respectively, $1-2^{-k^{+}_u}$ and $1$ in the first case and $2^{-k^{+}_u}$ and $0$ for the remaining cases. 
The average canalization depth reaches a maximum at the point $(r, s) = (0.5, 1)$, where the rule bias takes an intermediate value. 
For fixed values of $r$ and $s$, the algorithm requires the specification of an ordering of inputs. 
We consider two limiting cases:
\textit{concentrated control}, in which an incident node $v$ with the largest out-degree $k^-_v$ tends to canalize the other inputs, and \textit{diffuse control}, in which a node with the smallest $k^-_v$ tends to canalize the others. 
The former scenario is analogous to the disassortativity observed in the structure of regulatory networks, in which nodes with large $k^{-}$ tend to be connected to nodes with small $k^{-}$~\cite{Alexander2021}.
This situation allows for the cell's transcriptional state to be broadly altered by changing a select few transcription factors, sometimes referred to as ``general transcription factors'' in bacteria~\cite{Shen-Orr2002} or ``master regulators'' in eukaryotes~\cite{Lee2002}.
Under diffuse control on the other hand, genes referred to as ``specific transcription factors'' in bacteria~\cite{Shen-Orr2002} or ``secondary regulators’’ in eukaryotes~\cite{Lee2002} tend to canalize the output.
This scenario distributes control of the transcriptional state across many small circuits, enabling the spatiotemporal encoding of specific responses to particular signals~\cite{Boyer2005}.
Together, concentrated and diffuse control reflect competing strategies responsible for the organization of gene regulatory networks, with the latter case expected to have many more attractors than the former case. 
In \cref{fig:num_attractors}, we indeed observe a significantly larger number of attractors associated with networks in the latter scenario for parameters with large average canalization depth. 
The geometric mean over realizations ranges from
the order of $10^2$ attractors for concentrated control (descending sorting) to $10^3$ attractors for diffuse control (ascending sorting).

This framework allows us to probe the irreversibility in the core network of the $phoB$ origon $G'$.
We generate update rules $\mathbf{B}$ of the core network
for each $r = [0, 0.2, 0.4, 0.6, 0.8, 1]$ with $s=1$, and for each $s = [0, 0.2, 0.4, 0.6, 0.8, 1]$ with $r=1-s$ and $0$. 
For the $(r,s)$ pairs with nonunique rules, we sample $M=20$ realizations in each input order (for later reference, we define $M=1$ for unique rules). 
We determine the attractors for each realization of $\mathbf{B}$ using
an SAT-based algorithm~\cite{dubrova2011sat},
finding that the number of attractors varies with the rule nestedness and input sorting. 
The number of attractors is largest for intermediate values of the rule nestedness---though we note that the biological relevance of a given attractor varies widely.

\begin{figure*}[bt]
\centering
\includegraphics[width=160mm]{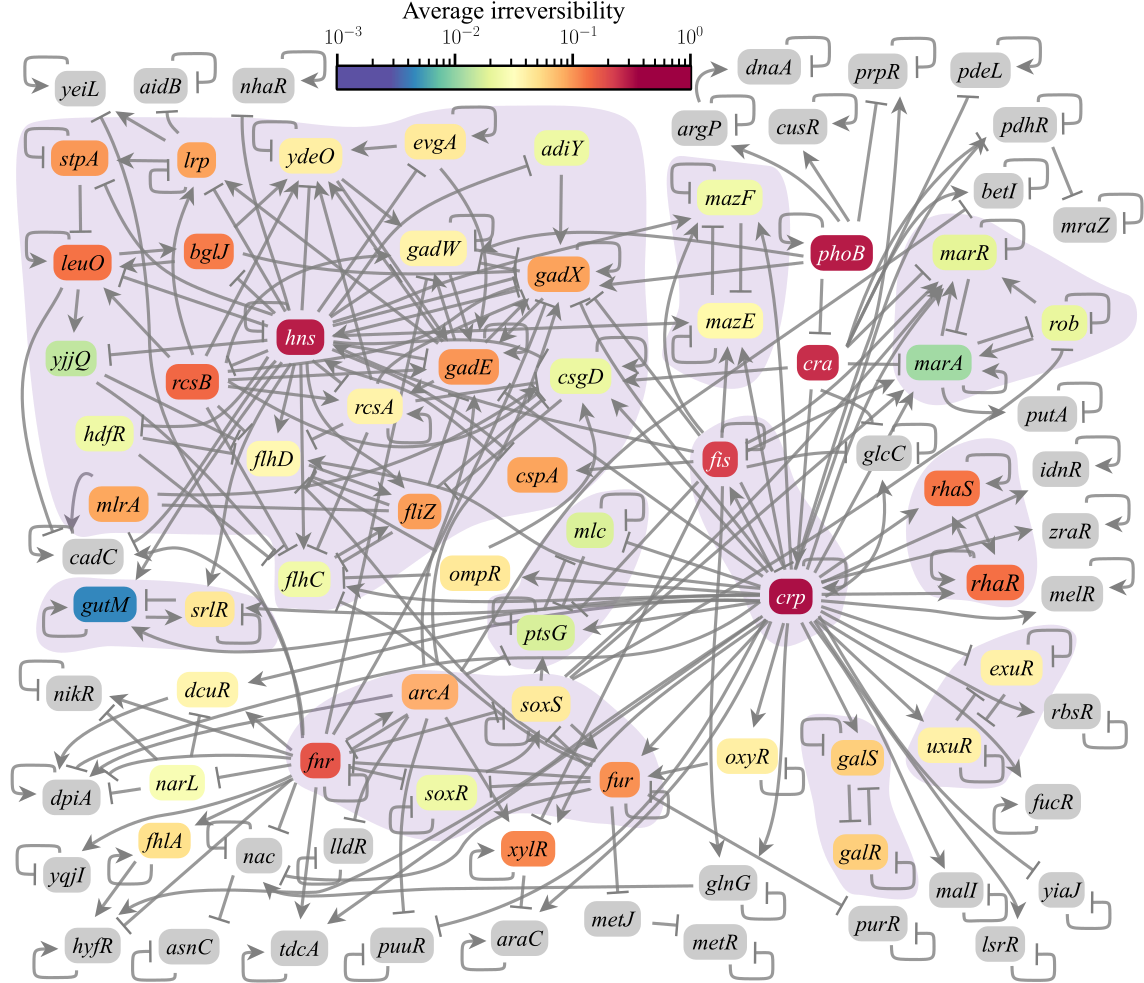}
\caption{\textbf{The 87-gene core regulatory network of the \textit{phoB} origon in \emph{E. coli}}. Node colors encode the average irreversibility probability across $M$ realizations of the rules, and edges denote regulatory interactions.
The shaded background indicates the multi-node SCCs, which all have one or more positive circuits. In total, 51 genes admit irreversible perturbations.
The average irreversibility probability across realizations is within 0.1 of the true value (\cref{fig:rules_ensemble}).
}
\vspace{-0.5cm}
\label{fig:result2}
\end{figure*}

\paragraph*{Analysis of the irreversibility results.} 
\Cref{fig:result2} summarizes the average probability that each gene in the \textit{phoB} origon core network admits an irreversible perturbation. 
Because such perturbations are on nodes that cause others to change when perturbed, these nodes reside in, or upstream of, nontrivial SCCs that contain at least one circuit with positive polarity. (An SCC is by definition a subnetwork for which each node can be reached by every other node, and we define trivial SCCs as single-node SCCs with no autoregulation.)
Circuits are directed loops in the network formed by a set of $m$ distinct nodes and $m$ distinct edges, and the circuit polarity is the product of the polarities of the edges.
Positive circuits are necessary for the existence of multiple fixed-point attractors~\cite{Remy2008}, which constitute the most common attractor class observed in our simulations. 
This creates the possibility of multiple stable states~\cite{Angeli2004,Craciun2011},
which we show below is a necessary condition for irreversibility.
The probability that the change is irreversible increases with proximity to downstream SCCs (shaded subnetworks in \cref{fig:result2}). 
Because large SCCs
tend to have more positive circuits, they create more opportunities for multistability,
which helps explain their observed proximity to upstream nodes admitting irreversible perturbations.
At the same time, the greater complexity of large SCCs means that they likely contain both positive and negative circuits. \Cref{fig:mixed_v_pos} shows that, in a minority of cases, this combination strengthens the irreversible response of selected genes compared to (smaller) SCCs with purely positive circuits by facilitating irreversible responses.
These results are an example of the network structure playing a role 
in determining irreversibility.

Every node influencing a positive circuit in the network exhibits irreversibility for some $\mathbf{B}$ in our simulations.
The remaining nodes cannot permanently alter the state of any positive circuit when transiently perturbed, and they are one of two types:
(i) leaf nodes (i.e., nodes with no outgoing edges to different nodes), which are reversible because they cannot influence other nodes; 
and (ii) nodes influencing only autorepressive leaf nodes, which
are reversible because the leaf node circuits are necessarily monostable.
We refer to the \nameref{sec:SM} for details and gene identities in each case.
Leaf nodes can still be irreversible response genes when they are downstream of an irreversibly responding positive circuit, which illustrates that
network structure also constrains the possible irreversible response genes.

To establish necessary and sufficient conditions for irreversibility, we examine the transitions induced by the application and removal of perturbations.
For a perturbation node $u$, we define the state inversion operator $g_u(\mathbf{x}) = \mathbf{x}|_{x_u=\bar{x}_u}$, and we refer to the time points in \cref{fig:illu}E.
The operator $g_u$ changes the state of variable $x_u$ to its inversion $\bar{x}_u$, while leaving the remaining states in $\mathbf{x}$ unchanged.
To prove the necessary condition by contradiction, suppose that the perturbation of $u$ is irreversible (i.e., $\mathcal{A}'\neq \mathcal{A}$) and that $g_u(\mathbf{x}^{Q})=\mathbf{x}^O$. 
But by definition $g_u(\mathbf{x}^{Q})=\mathbf{x}^{R}$, so $\left\{\mathbf{x}^O, \mathbf{x}^R, \mathbf{x}^{S}\right\} \subseteq \mathcal{A}$, making the perturbation reversible, a contradiction. 
As a consequence, there exists a nonempty set $\mathcal{W} = \{w\, |\, w \neq u,\, x^Q_w \neq x^O_w \}$.
Irreversibility further requires the existence of some $v \in V'$ such that
$B_v(\mathbf{x}^R) \neq B_v\big(g_{\mathcal{W}}(\mathbf{x}^R)\big)$, where we extend the $g$ operator to sets of nodes $\mathcal{W}$.
In the absence of such a $v$,
$\mathbf{x}^R$ and $g_{\mathcal{W}}(\mathbf{x}^R)$ belong to the same $\mathcal{A}'$, and $\mathcal{A}' = \mathcal{A}$ because $g_{\mathcal{W}}(\mathbf{x}^R) = \mathbf{x}^O$ by the definition of $\mathcal{W}$.
We can now state a sufficient condition for irreversibility in terms of the set $\mathcal{W}$ and the basin of attraction of $\mathcal{A}$, which is the set of states that reach $\mathcal{A}$ for some $t\geq 0$.
The condition is that $\mathbf{x}^R$ cannot be in the basin of attraction of $\mathcal{A}$, which implies $B_v(\mathbf{x}^R) \neq B_v\big(g_{\mathcal{W}}(\mathbf{x}^R)\big)$ for some $v \in V'$. 
Direct inspection of our simulations confirm that these conditions are satisfied in \cref{fig:result2}.

Since the basins of attraction are relevant to the sufficiency condition for irreversibility, we recalculated a weighted average of irreversibility in which the irreversibility in each initial attractor is weighted proportional to the size of its basin (\cref{fig:basinwt_analysis}).  
These weighted irreversibility results remain qualitatively similar to unweighted case, as indicated by the $R^2 > 0.91$ for both diffuse and concentrated control input orderings. However, the rate of irreversibility is cut approximately in half and the basin sizes of the initial attractors are on average approximately one-eighth the size of the final attractor basins. 
This tendency for irreversible perturbations to drive the network from attractors with smaller basins to those with larger basins can be understood as a consequence of the necessary and sufficient conditions: A set of downstream genes must change state in response to the irreversible perturbation (necessary condition), and the state reached upon reversion---provided that it is outside the original basin (sufficient condition)---is more likely to belong to a larger basin of attraction than a smaller one.

\begin{figure}[bt]\centering
\includegraphics[width=\textwidth]{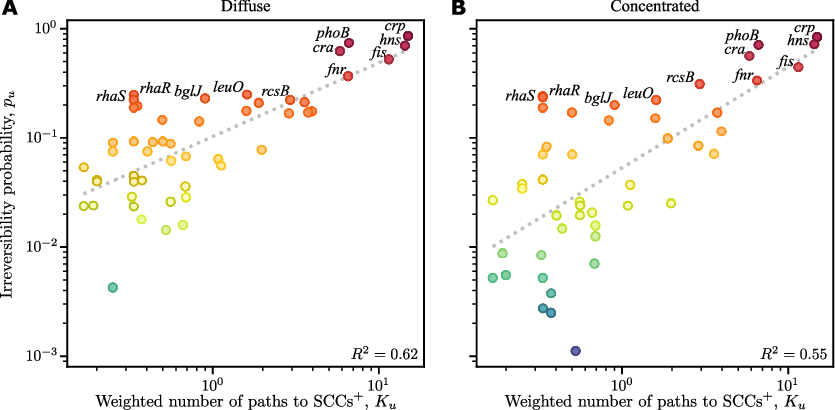}
\vspace{-0.25cm}
\caption{\textbf{Irreversibility probability for all nodes in the \textit{phoB} origon core network as a function of the weighted number of paths to SCCs with positive circuits ($\mathrm{SCCs}^+$).}
(\textbf{A} and \textbf{B}) Results for diffuse and concentrated control scenarios (i.e., ascending and descending input sorting), respectively.
The best fit trend is indicated by the dotted line in each case with the indicated coefficient of determination ($R^2$). The color code is the same as in \cref{fig:result2}.}
\label{fig:result3}
\vspace{-0.25cm}
\end{figure}

\paragraph*{Dynamical versus structural factors influencing irreversibility.} 
Casting irreversibility in terms of the set $\mathcal{W}$ allows us to relate $p_u$, the average irreversibility probability across all rules of node $u$ (a dynamical property), to the weighted number of paths to downstream SCCs with a positive circuit (a structural property). 
In \cref{fig:bool}C, we illustrate a simple network with two paths to SCCs with a positive circuit.
Each path is defined as a sequence of $\ell \geq 2$ nodes $\mathbf{H}^m=(H^m_1,...,H^m_\ell)$ (indexed by $m$), in which $(H^m_i, H^m_j) \in E'$ and $H^m_i \neq H^m_j$ for all $i \neq j$; in addition, $H^m_\ell \in \mathcal{K}_u$, where $\mathcal{K}_u$ is the set of nodes in all downstream SCCs of $G'$ with a positive circuit.
We argue that, starting at $u = H_1^m$, each path to $\mathcal{K}_u$ can contribute to the possibility that a perturbation 
gives rise to a nonempty $\mathcal{W}$.
We define the weight of path $\mathbf{H}^m$ to be
$\omega(\mathbf{H}^m)=(\prod_{i=2}^{\ell} k^+_{H^m_i})^{-1}$.
 (Example calculations of the path weight are provided in \cref{fig:bool}C.) Under certain approximations, the path weight corresponds to
the probability that the perturbation of $u$ changes the state of nodes in $\mathcal{K}_u$
through the path $\mathbf{H}^m$.
These approximations are that each input of $B_{H^m_i}$ is equally probable to change its output and that 
the change in state of each node in the path is independent.
Now, considering all paths from a node $u$ to $\mathcal{K}_u$, 
the weighted number of paths to all (other) nodes in strongly connected components for node $u$ is $K_u = \sum_{\left\{\mathbf{H}^m\right\}} \omega(\mathbf{H}^m)$.
\Cref{fig:result3} shows that 55--62\% of the variance in $p_u$ in our simulations is accounted for by the relationship $\hat{p}_u = a K_u^b$, where $b=0.68\pm0.09$ for diffuse control and $b=0.93\pm0.14$ for the concentrated control.
The larger exponent in the latter case reflects the 
reduced probability among nodes $u$ with a smaller number of weighted paths to $\mathcal{K}_u$.

\begin{figure*}[bt]\centering
\includegraphics[width=172mm]{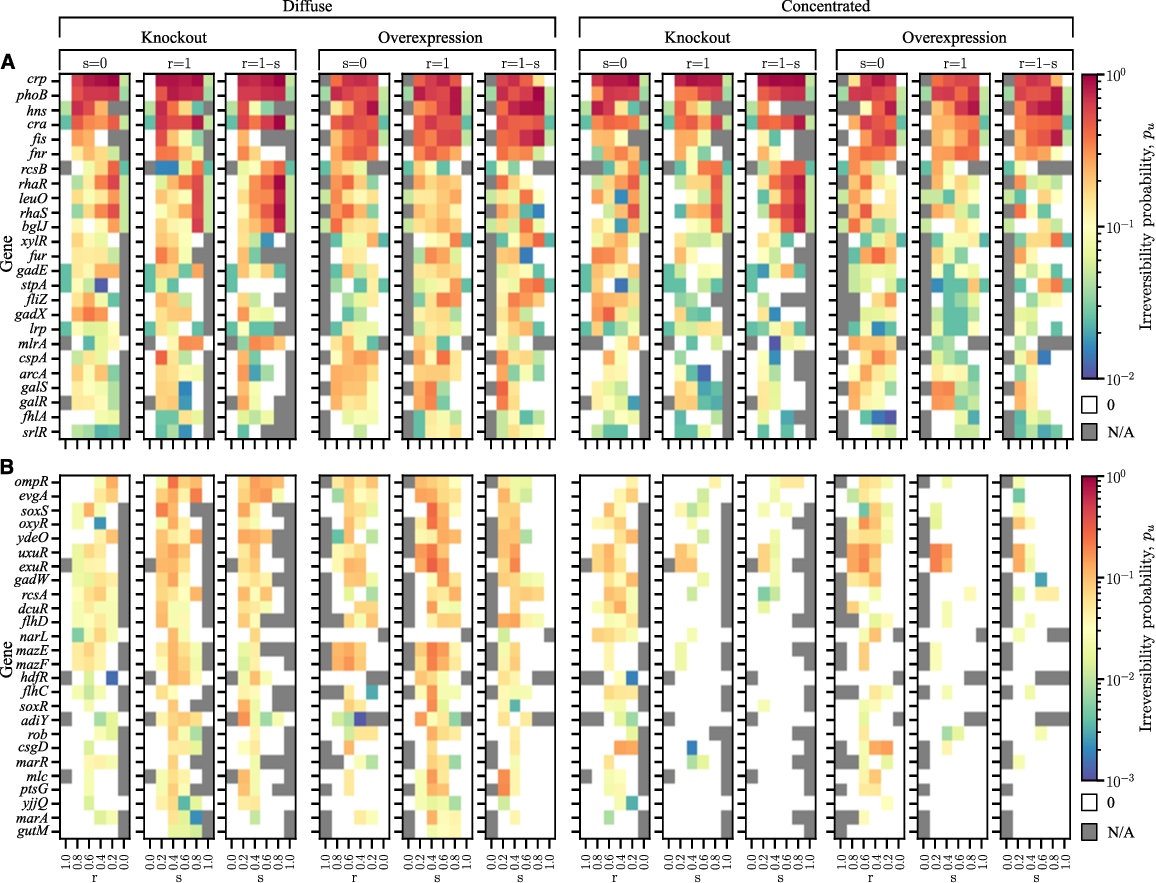}
\caption{\textbf{Probability of admitting irreversible perturbations averaged over realizations for the input orderings and perturbation types indicated above the panels.} 
(\textbf{A}) Color-coded irreversibility probability
as a function of $r$ and $s$ for the top 25 genes with the highest probability. 
For each input ordering and perturbation type, the three panels (from left to right) show the irreversibility probability along three different straight lines in the $(r,s)$-space in \cref{fig:rule-contours}. 
(\textbf{B}) Same plot as (A) for the remaining 26 genes admitting irreversible perturbations. The first and last column in each panel correspond to the cases of all inputs joined by $+$ and all inputs joined by $\times$, respectively.
}
\label{fig:all_genes_irrev}
\end{figure*}

\Cref{fig:all_genes_irrev} illustrates the trend in irreversibility averaged across realizations as a function of the input orderings, perturbations types, and values of $r$ and $s$ for the 25 most irreversible genes (\cref{fig:all_genes_irrev}A) and the remaining genes (\cref{fig:all_genes_irrev}B).
Within each panel varying $s$ and/or $r$, the columns are ordered such that the rule bias increases from left to right. 
The first and last columns are common to all panels of a given perturbation type because the rules are unique for these parameter choices.
In \cref{fig:all_genes_irrev}A, the irreversibility of the set of genes formed by \textit{hns}, \textit{stpA}, \textit{crp}, \textit{rcsB}, \textit{leuO}, \textit{bglJ}, \textit{rhaR}, and \textit{rhaS} varies monotonically with the rule bias in all three panels for each choice of ordering and perturbation type. 
Specifically, from left to right, the irreversibility decreases for the KO of \textit{hns} and the KO of \textit{stpA} 
but it increases for the KOs of the remaining genes in this set.
For each of these genes, as a function of the rule bias, the irreversibility of their OEs is anticorrelated with the irreversibility of their KOs. 
The anticorrelation is related to the number of attractors in which a particular gene is on (or off) in a given realization of the rules; that is, a perturbed gene tends to show greater irreversibility with respect to the perturbation (KO or OE) that can be applied to the largest number of attractors.
This can be intuitively understood by considering the limiting case in which only one attractor has a given gene on.
In this case, any irreversible transition induced by the KO of this gene requires the perturbed gene to change state \textit{after} the restoration of the KO. 
Such a change in state can only occur if the perturbed gene is in a circuit with other genes that change state after the initial perturbation.
This is in sharp contrast with the other extreme in which the given gene is on in all attractors, and thus irreversibility is possible even if it remains unchanged after the KO is restored. 

\Cref{fig:all_genes_irrev}A also shows that genes \textit{phoB}, \textit{cra}, and \textit{fis} exhibit a decrease in the irreversibility probability 
for intermediate values of $r$ and/or $s$,  which may be attributed to the larger average canalization depth for these parameter values. 
These genes have a large number of regulatory outputs $k^-_u$ and, due to the increased canalization depth, become less likely to determine the state updates compared to genes like \textit{fnr}, \textit{fur}, \textit{fliZ}, and \textit{gadX}, which have a lower overall irreversibility probability but show an increase in this probability for intermediate values of $r$ and/or $s$. 
Genes in the latter group tend to be located within large SCCs, while genes in the former group tend to be situated upstream of multiple SCCs. Finally, \cref{fig:all_genes_irrev}B shows the remaining 26 genes that admit irreversible perturbations but have smaller average irreversibility probability.
These genes tend to have a small out-degree ($k^-_u$), and thus they are most strongly affected by the input ordering.
In the descending input ordering, the genes with large out-degree dominate those with small  out-degree, resulting in the dearth of irreversibility for this ordering compared to the ascending ordering.

\begin{figure}[t!]
    \centering
    \includegraphics[width=0.6\textwidth]{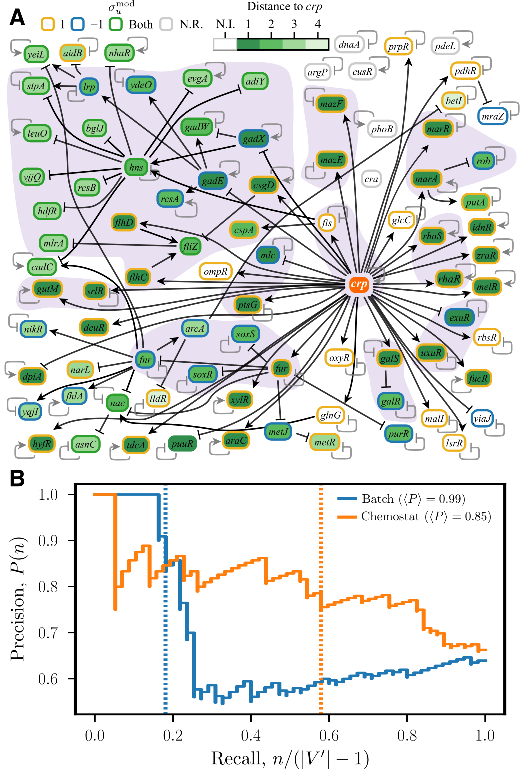}
    \caption{\textbf{Comparison of the irreversibility results with the observed transcriptional changes in adaptive evolution.}
    (\textbf{A}) Representation of the \textit{phoB} origon core network showing in black the edges that appear in the shortest paths to each node from \textit{crp}. The node outline colors indicate the sign of \textit{crp} regulation and the node colors indicate the distance from \textit{crp}, where irreversible response nodes are green. The shaded backgrounds, autoregulatory edges, and network layout are the same as in \cref{fig:result2}.
    (\textbf{B}) Precision-recall curves evaluating the agreement of the sign of expression change  of each gene predicted by the Boolean network model with that observed after adaptive evolution in batch (blue) and chemostat (orange) conditions. The genes are ranked from largest to smallest in terms of their change in expression. The dotted lines indicate the threshold of 0.5 for the log fold change used to calculate the $\langle P \rangle$ for each condition (marked on the legend). Genes above this threshold in batch and chemostat conditions are listed in \cref{tab:batch_adapt} and \cref{tab:chemo_adapt}, respectively. Abbreviations: N.R. -- Not Regulated, N.I. -- Not Irreversible.}
    \label{fig:prec_rec}
\end{figure}

\paragraph*{Irreversible genes in adaptive responses to \textit{crp} KO.}
The irreversible responses to transient genetic perturbations predicted here have implications for adaptive evolution. 
Intuitively, the response of the other genes to a gene KO followed by adaptive evolution is akin to the response to a gene KO followed by its reversion---in the sense that
both adaptation and the response to reversion tend to compensate for the changes induced by the initial perturbation.
In \cref{fig:prec_rec}, we examine this proposition by comparing the behavior of the 
genes in the core network that respond irreversibly to \textit{crp} KO---the most irreversible perturbation in our simulations---with those that do not in terms of their transcriptional changes during adaptive evolution to this perturbation. 
The gene \textit{crp} encodes the catabolite repressor protein, a global transcriptional regulator that represses genes associated with  the metabolism of non-preferred carbon sources in the presence of glucose. 
We make use of existing RNA-seq data from experiments where the cells were evolved for 10 days in M9 glucose
following \textit{crp} KO, 
which provides the highest-quality characterization of the transcriptome under these conditions~\cite{Pal2022}.
Using these data, we compute the observed sign and magnitude of the log fold change in expression between the initial and adaptively evolved strains, which were characterized under both batch and chemostat cultivation (details in \nameref{sec:methods}).
The observed sign for gene $u$, denoted $\sigma^{\rm obs}_u$, is compared against the sign predicted by the Boolean model $\sigma^{\rm mod}_u$. The latter is the \textit{opposite} of the polarity of the shortest path of \textit{crp} to the gene (\cref{fig:prec_rec}A).
Using $u'$ to denote the genes ordered in terms of decreasing magnitude of their log fold change, 
we compute the \textit{precision} of the top $n$ genes
\begin{equation}
    P(n) = \frac{1}{n} \sum_{u'=1}^{n} \mathbbm{1}\big(\sigma^{\rm obs}_{u'}=\sigma^{\rm mod}_{u'}\big), \label{eq:precision}
\end{equation}
which is the rate at which the signs match among these genes. 
Here, $\mathbbm{1}$ is the indicator function, which takes the value $1$ if its argument is true and $0$ if it is false. 
When there are multiple shortest paths of the same length to a given gene and two of these paths have different polarities, the indicator function evaluates to $1$ for nonzero values of $\sigma^{\rm obs}_{u}$.
\Cref{fig:prec_rec}B shows the precision $P(n)$ for both batch and chemostat conditions plotted as a function of $n$ (normalized by $|V'|-1$, the total number of genes in the \textit{phoB} origon core network other than \textit{crp}).
The precision decreases rapidly in both conditions when log fold change becomes less than a threshold of 0.5. 
Of the genes with a log fold change above this threshold, 10 of 11 in batch cultivation and 33 of 42 in chemostat cultivation change in the direction prescribed by the Boolean model.
Then, the average precision, defined by
\begin{equation}
    \langle P \rangle (n) = \frac{1}{n}\sum_{m=1}^n P(m), \label{eq:avg_precision} 
\end{equation}
is 0.99 and 0.85 in batch and chemostat cultivation, respectively (\cref{tab:batch_adapt,tab:chemo_adapt}). Both scenarios yield a significant $p$-value less than 0.01, as determined by bootstrapping (see \nameref{sec:methods}).

Having established that the correspondence between 
$\sigma^{\rm obs}_{u}$ and $\sigma^{\rm mod}_{u}$ is statistically significant, we examine the extent to which the genes with $\sigma^{\rm obs}_{u}=\sigma^{\rm mod}_{u}$ and a log fold change $> 0.5$ also corresponded to the 68 irreversible response genes associated with \textit{crp} KO in the Boolean model. 
We find that 8 of 9 irreversible response genes match the predicted response compared to 2 of 2 reversible genes in batch culture and 28 of 34 irreversible response genes match response compared to 5 of 8 reversible genes in chemostat culture, yielding  a $p$-value of 0.03 when considering both conditions together (see \nameref{sec:methods}). It is notable that a statistically significant relationship between the predicted irreversibility and adaptive evolution experiments is detected in spite of the limited information on the actual regulatory rules in the Boolean model and the non-regulatory factors known to influence adaptive evolution. 

\paragraph*{Making specific predictions}
Motivated by the concordance between the gene expression changes during adaptive evolution, we examine the irreversible perturbation of \textit{crp} KO in the context of existing transcriptional data and more detailed models of gene regulation (\cref{fig:scope_schematic}). 
First, we calculate whether each gene responds irreversibly to \textit{crp} KO across all attractors for all realizations of the rules.
We assess the biological relevance of the attractors by weighting the irreversibility results based on the
similarity of each attractor to the observed transcriptional states when calculating the average irreversibility (\cref{fig:wt_attr_analysis}).
This analysis leads us to conclude that self-activating genes that are positively regulated by \textit{crp} are the most likely to be irreversible.

While the level of detail in the Boolean model allows us to determine the type of perturbation (KO) and the initial states of the genes (both ON), it does not provide us with information regarding the continuous dynamics of the gene expression.
We obtain a continuous version of the Boolean dynamics that preserves the stable states by employing the HillCube algorithm~\cite{Wittmann2009} to represent the Boolean AND regulation as a differential equation.
In \cref{fig:diffeq_simulations}A, we use this representation to calculate the conditions for multistability in terms of phenomenological parameters like the transcriptional activation strength and Hill coefficient (a measure of how step-like the activation rule is).  
The irreversibility predictions from this analysis are verified by simulating the equations~(\cref{fig:diffeq_simulations}B).
From these simulations, we infer qualitative aspects that enhance irreversibility in this motif: irreversible response genes will tend to have stronger self-activation and exhibit a more switch-like response (i.e, have a larger Hill coefficient).

Overall, this analysis suggests candidate irreversible response genes such as \textit{zraR}, \textit{melR}, and \textit{rhaRS} in response to \textit{crp} KO.  
These genes are convenient because they one can ensure that they are initially on by adjusting the cultivation conditions (e.g., by using glycerol which is known to activate \textit{crp} and by supplementing the growth media metabolites such as zinc, melibiose, or rhamnose in the cases of \textit{zraR}, \textit{melR}, and \textit{rhaRS}, respectively).
Meanwhile, one candidate method for implementing the transient KO is inducible CRISPR interference~\cite{Qi2013}.
Finally, expression of the response genes could be monitored via sequencing to ascertain whether irreversibility occurs. 
Comparing with experiments exploring bistability in inducible sugar utilization~\cite{Afroz2014}, we posit that there will be a range of modest concentrations of supplemental metabolites the irreversible response gene will turn off, corresponding to the bistable region in~\cref{fig:diffeq_simulations}A.

\section*{Discussion}

The irreversibility of transient gene regulatory perturbations predicted here reveals a mechanism for prokaryotic cells to exhibit distinct phenotypes even when they are genetically identical. 
Our analysis, which excludes extracellular factors and chromatin modifications from the model, emphasizes the ability of purely regulatory mechanisms to precipitate 
heritable nongenetic changes that can endure for multiple generations. 
This should be compared with the phenomenon of cell fate commitment in eukaryotes, which is typically attributed to an environmental factor or signaling molecule triggering the expression of a master regulator that orchestrates the activation and repression of downstream genes to achieve a change in phenotype~\cite{Chang2008,Schultz2009}. 
In eukaryotes, epigenetic modifications such as histone modification and DNA methylation play a role in locking cells into their fates \cite{Sasai2013,Tripathi2020b}, but the former process is absent and the latter functions differently in prokaryotes. 

Within the scope of our \textit{E. coli} model, 
we establish 
that genes admitting irreversible perturbations
rely on positive circuits to generate multistability and stabilize their irreversible responses.
This finding reveals greatly enhanced complexity in the repertoire of possible cell states, well beyond those previously observed for bistable chemosensory motifs~\cite{Ozbudak2004,Afroz2014,Prajapat2015}.
Taken together, the results 
lead to the interesting possibility of nongenetically programming the state of bacterial cells---a phenomenon ultimately related to the control aim of steering between attractors in the regulatory network~\cite{Wells2015,Zanudo2017,Sullivan2023}.
Broadening our model to account for stochastic fluctuations, cell cycle, and other nonstationary factors can convert the permanently irreversible
responses seen in our model into temporarily irreversible but long-lived changes that persist over multiple generations. 
The extent to which the predicted irreversibility will persist
and be inheritable is thus an important question for future
experimental studies, which 
can be interpreted using stochastic many-body physics approaches tailored to describe the processes of transcription, translation, and degradation~\cite{Hornos2005,Walczak2005,Tkacik2012,Assaf2013,Zhang2013,Bhattacharyya2020}.

Finally, our analysis suggests that genes responding irreversibly are significantly associated with those that undergo regulatory changes in adaptive evolution experiments across conditions, even in the absence of full knowledge of the regulatory rules.
This result is consistent with the observation that incomplete models of gene regulatory networks can still yield reliable predictions~\cite{Tripathi2023}.
Thus, notwithstanding the simplifications of the model, the analysis of irreversible responses to \textit{transient} perturbations also contributes to the interpretation of adaptive evolution responses to \textit{permanent} perturbations.

\section*{Materials and Methods}
\setcurrentname{Materials and Methods}
\label{sec:methods}

\paragraph*{Construction of the \textit{phoB} origon core network.} 
We constructed the activating and repressing interactions of the gene regulatory network based on the RegulonDB data using the file ``generegulation\_tmp.txt''~\cite{Santos-Zavaleta2018}, where pairwise  regulatory relationships between genes are recorded. 
The regulatory network dynamics were simulated using R (version 4.2.3) and R package BoolNet (version 2.1.8)~\cite{mussel2010boolnet}.
In order for the dynamics to be well defined,
each node must have at least one input, where we recognize the rule $x^{t+1}_u=x^{t}_u$ as positive autoregulation.
Therefore, in analyzing the \textit{phoB} origon, we added a self-activating loop  (and no additional inputs) to \textit{phoB} 
as this is the only gene in the core network with no regulatory inputs. 
This added edge fixes the initial state of the node and does not impact our observation of irreversibility.

\paragraph*{Generation of biologically realistic update rules.}
Because the available RegulonDB data are insufficient to specify all the Boolean update rules,
we examine the ensemble of consistent rules.
Rules are said to be \textit{consistent} with RegulonDB if they satisfy the three criteria laid out in the main text:
\begin{enumerate}
    \item[1.] \textit{Edge consistency.} The state variables $x_v$ (or their negation $\bar{x}_v$)
appearing on the right-hand side of \cref{eq:Bool} are those associated with nodes $v$ that have edges incident on $u$ in $G'$. 
    \item[2.] \textit{Edge essentiality.} 
Whenever $v$ is a node incident on $u$ in $G'$, there is at least one state $\mathbf{x}$ for which changing the variable $x_v$ changes  $B_u(\mathbf{x})$.
    \item[3.] \textit{Sign consistency.} 
We require that $B_u(\mathbf{x}|_{x_v=0}) \leq B_u(\mathbf{x}|_{x_v=1})$ if $v$ activates $u$ and $B_u(\mathbf{x}|_{x_v=0}) \geq B_u(\mathbf{x}|_{x_v=1})$ if $v$ inhibits $u$.

\end{enumerate}
To exclude artifactual oscillations, we further assume that autorepressive regulation is silenced
when $x_u=0$ in the sign consistency condition, which implies an exception to the edge essentiality condition. 
Specifically for every rule chosen, one or more edges into a autorepressive node will not influence the state of this node. 
If the autorepressive node is in its own monomial, the self-edge is nonessential.
If the autorepressive node is joined with others in a monomial, then the other input edges to this node in the monomial will be nonessential.

\paragraph*{Estimating the number of possible regulatory rules.}
Edge and sign consistency together guarantee that one of $x^t_v$ or $\bar{x}^t_v$ appear in the rule $B_u$, and edge essentiality guarantees that all $v$ must appear once. Thus, the number of  variables in the rule $B_u$ is $k_u^+$.
Since all possible Boolean rules can be written as a sum of products~\cite{he2016stratification},
the number of feasible rules is at least as large as $\sum_{n=0}^{k^+_u-1} \binom{k^+_u - 1}{n}=2^{k^+_u-1}$.
Because the $B_u$ are set independently for each $u$, the number of feasible $\mathbf{B}$ is $\prod_{u=1}^{|V'|} 2^{k_u^+-1}$, which of the order of $10^{61}$ for the \textit{phoB} origon core network.

\paragraph*{Algorithm to sample realistic update rules.}
In the general case, given a network, there is an \textit{ensemble} of possible $\mathbf{B}$ that are consistent with the network structure and polarity of the interactions.
To facilitate the sampling of this ensemble, we introduce the vector of inputs $\mathbf{y}=[y_{v_1},\ldots, y_{v_{k^+_u}}\ |\ (v_i, u) \in E'\,]$ for fixed $u$, where  $y_{v_i}=x_{v_i}$ if $W(v_i, u)>0$ and $y_{v_i}=\bar{x}_{v_i}$ if $W(v_i, u)<0$.
The inputs are indexed using $i \in \{1,\ldots, k^+_u\}$ and ordered according to $k^-_{v_i}$, the number of outgoing edges of the associated node $v_i$, for both ascending ($k^-_{v_i}\leq k^-_{v_{i+1}}$) and descending ($k^-_{v_i}\geq k^-_{v_{i+1}}$) orders.
The sampling of the ensemble of rules is parameterized by $r$ and $s$, 
which control the selection among three binary operators between inputs $y_{v_i}$ and $y_{v_{i+1}}$. 
Specifically, we join the inputs as
$y_{v_i} \times (y_{v_{i+1}}$ with probability $r$, as $y_{v_i} + y_{v_{i+1}}$ with probability $(1-r)s$, and as $y_{v_i} \times y_{v_{i+1}}$ with probability $(1-r)(1-s)$.
Larger values of the nestedness parameter increase the number of parentheses appearing in the rules, and larger values of the bias parameter increase the number of possible input vectors that update to 1.
We specify each $B_u$ by starting at the first pair of inputs, choosing their binary operator according to the probabilities above, and proceeding iteratively until all inputs are included.
This strategy is implemented in the following iterative algorithm:

\begin{algorithmic}
\baselineskip12pt
\Procedure{Generate rule$\,$}{$r, s, u, k^+_u,\mathbf{y}$}
   \State $n\gets 0$
   \State $i\gets 1$
   \State $B_u^{(i)} \gets y_{v_i}$
   \While{$i<k^{+}_{u}$}\Comment{Iterate over all inputs}
      \State $\xi_1 \sim \mathbbm{U}(0,1)$
      \If{$\xi_1 < r$}
        \State $B_u^{(i+1)} \gets B_u^{(i)}\ \times\ ( y_{v_{i+1}}$ \Comment{Open parentheses}
        \State $n\gets n+1$
      \Else
        \State $\xi_2 \sim \mathbbm{U}(0,1)$
        \If{$\xi_2 < s$}
          \State $B_u^{(i+1)} \gets B_u^{(i)}\ + \ y_{v_{i+1}}$
        \Else
          \State $B_u^{(i+1)} \gets B_u^{(i)}\ \times\ y_{v_{i+1}}$
        \EndIf
      \EndIf
      \State $i\gets i+1$
   \EndWhile
   \State $B_u \gets B_u^{(k^+_u)})...)$ \Comment{Close matching parentheses}
   \State \textbf{return} $B_u$
\EndProcedure
\end{algorithmic}
\baselineskip24pt
In the algorithm, we use $\sim \mathbbm{U}(0,1)$ to denote a random number drawn from the uniform distribution on the unit interval, and we use $)...)$ to denote the $n$ closed parentheses in the rule.

\paragraph*{Convergence of irreversibility estimates.}
For each pair $(r,s)$ and input sorting that does not have unique rules [$r\neq1,\ s\neq0, $ and $(r,s)\neq(0,1)$], we generate $M=20$ realizations of the rules indexed by $i \in \{1,\ldots, M\}$,
identify the attractors (\cref{fig:illu}D),
and apply the irreversibility detection algorithm~(\cref{fig:illu}E).
Since the attractors change between realizations of the rules, the irreversibility of a perturbation (KO or OE) may also change. 
To account for this source of variability, we average over the transient perturbations as follows.
Let $q_{u, i}$ be the fraction of attractors for which $x_u=1$, and let $p_{u, i}^{\mathrm{KO}}$ and $p_{u, i}^{\mathrm{OE}}$ be the probabilities that the transient KO and OE of gene $u$ lead to irreversibility, respectively. 
Then, 
\begin{equation}
    \widehat{p}^{\,\,\mathcal{S}}_u = \frac{1}{|\mathcal{S}|} \sum_{i\in\mathcal{S}} p^{\mathrm{KO}}_{u, i} q_{u, i} +p^{\mathrm{OE}}_{u, i} (1-q_{u, i}), \label{eq:avg_over_pert}
\end{equation}
is the weighted average of the probability that gene $u$ admits an irreversible perturbation across a set of realizations $\mathcal{S}$. 

We test for the convergence of the average irreversibility as a function of the ensemble size by fixing the number of realizations to be $M'=\{1, ..., 10\}$.
The number of possible ensembles of size $M'$ taken out of $M$ realizations is $\binom{M}{M'}$, and the number of pairs of ensembles is $Z(M,M') = \binom{M}{M'}\binom{M-M'}{M'}$.
If $Z(M,M')>1{,}000$, we randomly sample 
$1{,}000$ pairs of ensembles. Otherwise, we use all $Z(M,M')$ pairs. 
Denoting each ensemble pair as $(\mathcal{U},\mathcal{V})$,
we apply \cref{eq:avg_over_pert} to each ensemble to obtain the root mean square difference
\begin{equation}
    \mathrm{RMSD} = \sqrt{\frac{1}{|V'|}\sum_{u=1}^{|V'|} \left(\widehat{p}^{\,\,\mathcal{U}}_u - \widehat{p}^{\,\,\mathcal{V}}_u\right)^2}, \label{eq:RMSD}
\end{equation}
where RMSD stands for Root Mean Square Deviation. We 
recall that $|V'|=87$ is the number of genes in the core network. 

\paragraph*{Processing of the RNA-seq data.}
The transcriptional data for \textit{E. coli} adaptively evolved after \textit{crp} KO are obtained and analyzed as follows.
Raw counts of RNA were obtained from the Gene Expression Omnibus (GEO) database~\cite{Barrett2013} maintained by the National Center for Biotechnology Information (NCBI), accession number GSE152214. 
Experimental 
details of the RNA collection have been described elsewhere~\cite{Pal2022}.
Raw counts in GEO were converted into transcripts per million (TPM) using 
 $z_i = 10^6 \frac{c_i}{L_i} \big(\sum_{j=1}^{N_g} \frac{c_j}{L_j} \big)^{-1}$, where $N_g$ is the number of genes in the dataset,  $c_i$ is the raw count of transcripts for gene $i$, and $L_i$ is the length of the gene in kilobases.
The transcript counts of genes that are in the core regulatory network $G'$ in our model were examined for changes before \emph{crp} KO, after \emph{crp} KO, and after adaptive evolution of the \textit{crp} KO strain. 
The data include strains cultivated under both batch and chemostat conditions. 

\paragraph*{Calculating the observed sign of the transcriptional changes.}
For each environmental condition, $\rho_u =  \mu_u^{\rm evo} / \mu_u^{\rm wt}$ is the fold change of the average expression for each gene between the adaptively evolved strain and the initial wild-type strain. 
To account for the overall shift in transcription (e.g., due to changes in the lab conditions or variability in media preparation), we calculate the average shift in expression $\langle \rho \rangle =   |V'|^{-1} \sum_{u=1}^{|V'|} \mu_u^{\rm evo} / \mu_u^{\rm wt}$.
Then, the observed sign of regulatory changes is 
\begin{equation}
    \sigma^{\rm obs}_u = \operatorname{sgn}\left(\ln \frac{\rho_u}{\langle \rho \rangle} \right), \label{eq:sign_obs}
\end{equation}
where $\operatorname{sgn}(\epsilon)$ is the sign function that takes the value $1$ if  $\epsilon>0$, $-1$ if $\epsilon<0$, and $0$ if  $\epsilon=0$. 
In addition, the magnitude of the log fold change in expression is $\big|\ln \big(\rho_u / \langle \rho \rangle\big)\big|$.

\paragraph*{Calculating the predicted sign of the transcriptional changes.}
The predicted sign of the transcriptional change in the Boolean model is assigned according to the polarity of the shortest paths in $G'$ from \textit{crp} to each gene $u$.
Recall that the a shortest path between $H^u_1 = \mbox{\textit{crp}}$ and $H^u_\ell = u$ is denoted by $\mathbf{H}^u = (H^u_1,...,H^u_\ell)$ and that the polarity of the edges is given by the function $W$. 
According to the Boolean model, the 
sign of the expected change is
\begin{equation}
    \sigma^{\rm mod}_u = - \prod_{i=1}^{\ell-1} W(H^u_i, H^u_{i+1}), \label{eq:sign_mod}
\end{equation}
where the negative sign appears because the perturbation of \textit{crp} is a KO.
\Cref{eq:sign_obs,eq:sign_mod} provide the quantities used in calculating the precision in \cref{eq:precision}.

\paragraph*{Statistical analysis of $\langle P \rangle$ and irreversibility.}
We assess statistical significance using a bootstrapping approach~\cite{Efron1985}.
In this approach, we compute $N_{\rm exc}$, the number of times that the randomized list returns a larger value of the statistic than the observed lists out of $N_{\rm samp}=25{,}000$ shufflings, and the $p$-value is given by $1-N_{\rm exc}/N_{\rm samp}$.
Specifically, we compute $\langle P \rangle$ in each condition to evaluate whether the concordance between $\sigma^{\rm obs}_u$ and $\sigma^{\rm mod}_u$ is statistically significant. 
Using $u'$ to denote the reordered list of genes, 
$\langle P \rangle$ is repeatedly computed after shuffling $\sigma^{\rm mod}_{u'}$ while keeping $\sigma^{\rm obs}_u$ and $\big|\ln \big(\rho_u / \langle \rho \rangle\big)\big|$ fixed. 

Similarly, we compute the number of genes that satisfy $\sigma^{\rm obs}_u=\sigma^{\rm mod}_u$ and $\big|\ln \big(\rho_u / \langle \rho \rangle\big)\big|>0.5$
for genes responding irreversibly ($\gamma_{\rm irr}$) and reversibly ($\gamma_{\rm rev}$) to \textit{crp} KO in each cultivation condition.
Using the difference $\gamma_{\rm irr}-\gamma_{\rm rev}$ in each condition, we assess whether genes that show large changes during adaptive evolution are significantly more likely to be irreversible in our simulations.
In this case, we repeatedly compute 
$\gamma_{\rm irr}-\gamma_{\rm rev}$
after shuffling $\sigma^{\rm mod}_{u'}$ while keeping $\sigma^{\rm obs}_u$ and $\big|\ln \big(\rho_u / \langle \rho \rangle\big)\big|$ fixed in each condition separately. We count cases toward $N_{\rm exc}$ only when
$\gamma_{\rm irr}-\gamma_{\rm rev}$ exceeds the observed value in both environmental conditions.

\bibliography{irreversibility}

\begin{thebibliography}{10}

\bibitem{Koch1994}
A.~J. Koch, H.~Meinhardt, {Biological pattern formation: {F}rom basic
  mechanisms to complex structures}.
\newblock {\it Rev. Mod. Phys.\/} {\bf 66}, 1481--1507 (1994).

\bibitem{Winfree2001}
A.~T. Winfree, {\it The Geometry of Biological Time\/} (Springer, New York,
  2001).

\bibitem{Lynn2022}
C.~W. Lynn, C.~M. Holmes, W.~Bialek, D.~J. Schwab, {Emergence of local
  irreversibility in complex interacting systems}.
\newblock {\it Phys. Rev. E\/} {\bf 106}, 034102 (2022).

\bibitem{Keim2019}
N.~C. Keim, J.~D. Paulsen, Z.~Zeravcic, S.~Sastry, S.~R. Nagel, {Memory
  formation in matter}.
\newblock {\it Rev. Mod. Phys.\/} {\bf 91}, 035002 (2019).

\bibitem{Crick1970}
F.~H.~C. Crick, {Central Dogma of Molecular Biology}.
\newblock {\it Nature\/} {\bf 227}, 561--563 (1970).

\bibitem{Villarreal2012}
C.~Villarreal, P.~Padilla-Longoria, E.~R. Alvarez-Buylla, {General Theory of
  Genotype to Phenotype Mapping: Derivation of Epigenetic Landscapes from
  N-Node Complex Gene Regulatory Networks}.
\newblock {\it Phys. Rev. Lett.\/} {\bf 109}, 118102 (2012).

\bibitem{Huang2005}
S.~Huang, G.~Eichler, Y.~Bar-Yam, D.~E. Ingber, {Cell Fates as High-Dimensional
  Attractor States of a Complex Gene Regulatory Network}.
\newblock {\it Phys. Rev. Lett.\/} {\bf 94}, 128701 (2005).

\bibitem{Zhou2011}
J.~X. Zhou, S.~Huang, {Understanding gene circuits at cell-fate branch points
  for rational cell reprogramming}.
\newblock {\it Trends Genet.\/} {\bf 27}, 55--62 (2011).

\bibitem{Zanudo2015}
J.~G.~T. Za{\~{n}}udo, R.~Albert, {Cell Fate Reprogramming by Control of
  Intracellular Network Dynamics}.
\newblock {\it PLOS Comput. Biol.\/} {\bf 11}, e1004193 (2015).

\bibitem{Zhu2022}
R.~Zhu, J.~M. del Rio-Salgado, J.~Garcia-Ojalvo, M.~B. Elowitz, Synthetic
  multistability in mammalian cells.
\newblock {\it Science\/} {\bf 375}, eabg9765 (2022).

\bibitem{Lopez-Otin2013}
C.~L{\'{o}}pez-Ot{\'{i}}n, M.~A. Blasco, L.~Partridge, M.~Serrano, G.~Kroemer,
  {The Hallmarks of Aging}.
\newblock {\it Cell\/} {\bf 153}, 1194--1217 (2013).

\bibitem{He2017}
S.~He, N.~E. Sharpless, {Senescence in Health and Disease}.
\newblock {\it Cell\/} {\bf 169}, 1000--1011 (2017).

\bibitem{Shapiro1971}
L.~Shapiro, N.~Agabian-Keshishian, I.~Bendis, {Bacterial Differentiation}.
\newblock {\it Science\/} {\bf 173}, 884--892 (1971).

\bibitem{Veening2008}
J.-W. Veening, W.~K. Smits, O.~P. Kuipers, {Bistability, Epigenetics, and
  Bet-Hedging in Bacteria}.
\newblock {\it Annu. Rev. Microbiol.\/} {\bf 62}, 193--210 (2008).

\bibitem{Ozbudak2004}
E.~M. Ozbudak, M.~Thattai, H.~N. Lim, B.~I. Shraiman, A.~van Oudenaarden,
  {Multistability in the lactose utilization network of \emph{Escherichia
  coli}}.
\newblock {\it Nature\/} {\bf 427}, 737--740 (2004).

\bibitem{Prajapat2015}
M.~K. Prajapat, K.~Jain, S.~Saini, {Control of MarRAB Operon in
  \emph{Escherichia coli} via Autoactivation and Autorepression}.
\newblock {\it Biophys. J.\/} {\bf 109}, 1497--1508 (2015).

\bibitem{Santos-Zavaleta2018}
A.~Santos-Zavaleta, M.~Sánchez-Pérez, H.~Salgado, D.~A. Velázquez-Ramírez,
  S.~Gama-Castro, V.~H. Tierrafría, S.~J.~W. Busby, P.~Aquino, X.~Fang, B.~O.
  Palsson, J.~E. Galagan, J.~Collado-Vides, A unified resource for
  transcriptional regulation in \emph{Escherichia coli} {K}-12 incorporating
  high-throughput-generated binding data into {RegulonDB} version 10.0.
\newblock {\it BMC Biol.\/} {\bf 16}, 91 (2018).

\bibitem{Balazsi2005}
G.~Balazsi, A.-L. Barabasi, Z.~N. Oltvai, {Topological units of environmental
  signal processing in the transcriptional regulatory network of
  \emph{Escherichia coli}}.
\newblock {\it Proc. Natl. Acad. Sci. USA\/} {\bf 102}, 7841--7846 (2005).

\bibitem{Samal2008}
A.~Samal, S.~Jain, {The regulatory network of \emph{E. coli} metabolism as a
  Boolean dynamical system exhibits both homeostasis and flexibility of
  response}.
\newblock {\it BMC Syst. Biol.\/} {\bf 2}, 21 (2008).

\bibitem{Azimi-Tafreshi2013}
N.~Azimi-Tafreshi, S.~N. Dorogovtsev, J.~F.~F. Mendes, {Core organization of
  directed complex networks}.
\newblock {\it Phys. Rev. E\/} {\bf 87}, 032815 (2013).

\bibitem{Kauffman1993}
S.~A. Kauffman, {\it The Origins of Order: Self-organization and Selection in
  Evolution\/} (Oxford University Press, New York, 1993).

\bibitem{Wang2012}
R.-S. Wang, A.~Saadatpour, R.~Albert, {Boolean modeling in systems biology: An
  overview of methodology and applications}.
\newblock {\it Phys. Biol.\/} {\bf 9}, 055001 (2012).

\bibitem{he2016stratification}
Q.~He, M.~Macauley, Stratification and enumeration of {B}oolean functions by
  canalizing depth.
\newblock {\it Phys. D\/} {\bf 314}, 1--8 (2016).

\bibitem{Harris2002}
S.~E. Harris, B.~K. Sawhill, A.~Wuensche, S.~Kauffman, {A model of
  transcriptional regulatory networks based on biases in the observed
  regulation rules}.
\newblock {\it Complexity\/} {\bf 7}, 23--40 (2002).

\bibitem{Kauffman2003}
S.~Kauffman, C.~Peterson, B.~Samuelsson, C.~Troein, {Random Boolean network
  models and the yeast transcriptional network}.
\newblock {\it Proc. Natl. Acad. Sci. USA\/} {\bf 100}, 14796--14799 (2003).

\bibitem{Moreira2005}
A.~A. Moreira, L.~A.~N. Amaral, {Canalizing Kauffman Networks: Nonergodicity
  and Its Effect on Their Critical Behavior}.
\newblock {\it Phys. Rev. Lett.\/} {\bf 94}, 218702 (2005).

\bibitem{Quine1955}
W.~V. Quine, {A Way to Simplify Truth Functions}.
\newblock {\it Am. Math. Mon.\/} {\bf 62}, 627 (1955).

\bibitem{McCluskey1956}
E.~J. McCluskey, {Minimization of Boolean Functions}.
\newblock {\it Bell Syst. Tech. J.\/} {\bf 35}, 1417--1444 (1956).

\bibitem{Pomerance2009}
A.~Pomerance, E.~Ott, M.~Girvan, W.~Losert, {The effect of network topology on
  the stability of discrete state models of genetic control}.
\newblock {\it Proc. Natl. Acad. Sci. USA\/} {\bf 106}, 8209--8214 (2009).

\bibitem{Squires2014}
S.~Squires, A.~Pomerance, M.~Girvan, E.~Ott, {Stability of Boolean networks:
  The joint effects of topology and update rules}.
\newblock {\it Phys. Rev. E\/} {\bf 90}, 022814 (2014).

\bibitem{Tripathi2020}
S.~Tripathi, D.~A. Kessler, H.~Levine, {Biological Networks Regulating Cell
  Fate Choice are Minimally Frustrated}.
\newblock {\it Phys. Rev. Lett.\/} {\bf 125}, 088101 (2020).

\bibitem{Tripathi2023}
S.~Tripathi, D.~A. Kessler, H.~Levine, {Minimal frustration underlies the
  usefulness of incomplete regulatory network models in biology}.
\newblock {\it Proc. Natl. Acad. Sci. USA\/} {\bf 120}, e2216109120 (2023).

\bibitem{Alexander2021}
B.~Alexander, A.~Pushkar, M.~Girvan, {Phase transitions and assortativity in
  models of gene regulatory networks evolved under different selection
  processes}.
\newblock {\it J. R. Soc. Interface\/} {\bf 18}, 20200790 (2021).

\bibitem{Shen-Orr2002}
S.~S. Shen-Orr, R.~Milo, S.~Mangan, U.~Alon, {Network motifs in the
  transcriptional regulation network of \textit{Escherichia coli}}.
\newblock {\it Nat. Genet.\/} {\bf 31}, 64--68 (2002).

\bibitem{Lee2002}
T.~I. Lee, N.~J. Rinaldi, F.~Robert, D.~T. Odom, Z.~Bar-Joseph, G.~K. Gerber,
  N.~M. Hannett, C.~T. Harbison, C.~M. Thompson, I.~Simon, J.~Zeitlinger, E.~G.
  Jennings, H.~L. Murray, D.~B. Gordon, B.~Ren, J.~J. Wyrick, J.-B. Tagne,
  T.~L. Volkert, E.~Fraenkel, D.~K. Gifford, R.~A. Young, {Transcriptional
  Regulatory Networks in \textit{Saccharomyces cerevisiae}}.
\newblock {\it Science\/} {\bf 298}, 799--804 (2002).

\bibitem{Boyer2005}
L.~A. Boyer, T.~I. Lee, M.~F. Cole, S.~E. Johnstone, S.~S. Levine, J.~P.
  Zucker, M.~G. Guenther, R.~M. Kumar, H.~L. Murray, R.~G. Jenner, D.~K.
  Gifford, D.~A. Melton, R.~Jaenisch, R.~A. Young, {Core Transcriptional
  Regulatory Circuitry in Human Embryonic Stem Cells}.
\newblock {\it Cell\/} {\bf 122}, 947--956 (2005).

\bibitem{dubrova2011sat}
E.~Dubrova, M.~Teslenko, A {SAT}-based algorithm for finding attractors in
  synchronous {B}oolean networks.
\newblock {\it IEEE/ACM Transactions on Computational Biology and
  Bioinformatics\/} {\bf 8}, 1393--1399 (2011).

\bibitem{Remy2008}
{\'{E}}.~Remy, P.~Ruet, D.~Thieffry, {Graphic requirements for multistability
  and attractive cycles in a Boolean dynamical framework}.
\newblock {\it Adv. Appl. Math.\/} {\bf 41}, 335--350 (2008).

\bibitem{Angeli2004}
D.~Angeli, J.~E. Ferrell, E.~D. Sontag, {Detection of multistability,
  bifurcations, and hysteresis in a large class of biological positive-feedback
  systems}.
\newblock {\it Proc. Natl. Acad. Sci. USA\/} {\bf 101}, 1822--1827 (2004).

\bibitem{Craciun2011}
G.~Craciun, C.~Pantea, E.~D. Sontag, {\it Design and Analysis of Biomolecular
  Circuits: Engineering Approaches to Systems and Synthetic Biology\/},
  H.~Koeppl, G.~Setti, M.~di~Bernardo, D.~Densmore, eds. (Springer New York,
  New York, NY, 2011), pp. 63--72.

\bibitem{Pal2022}
A.~Pal, M.~S. Iyer, S.~Srinivasan, A.~S. {Narain Seshasayee}, K.~V. Venkatesh,
  {Global pleiotropic effects in adaptively evolved \emph{Escherichia coli}
  lacking CRP reveal molecular mechanisms that define the growth physiology}.
\newblock {\it Open Biol.\/} {\bf 12}, 210206 (2022).

\bibitem{Wittmann2009}
D.~M. Wittmann, J.~Krumsiek, J.~Saez-Rodriguez, D.~A. Lauffenburger, S.~Klamt,
  F.~J. Theis, Transforming {Boolean} models to continuous models:
  {Methodology} and application to {T}-cell receptor signaling.
\newblock {\it BMC Syst. Biol.\/} {\bf 3}, 98 (2009).

\bibitem{Qi2013}
L.~S. Qi, M.~H. Larson, L.~A. Gilbert, J.~A. Doudna, J.~S. Weissman, A.~P.
  Arkin, W.~A. Lim, {Repurposing CRISPR as an RNA-Guided Platform for
  Sequence-Specific Control of Gene Expression}.
\newblock {\it Cell\/} {\bf 152}, 1173--1183 (2013).

\bibitem{Afroz2014}
T.~Afroz, K.~Biliouris, Y.~Kaznessis, C.~L. Beisel, {Bacterial sugar
  utilization gives rise to distinct single-cell behaviours}.
\newblock {\it Mol. Microbiol.\/} {\bf 93}, 1093--1103 (2014).

\bibitem{Chang2008}
H.~H. Chang, M.~Hemberg, M.~Barahona, D.~E. Ingber, S.~Huang,
  {Transcriptome-wide noise controls lineage choice in mammalian progenitor
  cells}.
\newblock {\it Nature\/} {\bf 453}, 544--547 (2008).

\bibitem{Schultz2009}
D.~Schultz, P.~G. Wolynes, E.~B. Jacob, J.~N. Onuchic, {Deciding fate in
  adverse times: Sporulation and competence in \emph{Bacillus subtilis}}.
\newblock {\it Proc. Natl. Acad. Sci. USA\/} {\bf 106}, 21027--21034 (2009).

\bibitem{Sasai2013}
M.~Sasai, Y.~Kawabata, K.~Makishi, K.~Itoh, T.~P. Terada, {Time Scales in
  Epigenetic Dynamics and Phenotypic Heterogeneity of Embryonic Stem Cells}.
\newblock {\it PLoS Comput. Biol.\/} {\bf 9}, e1003380 (2013).

\bibitem{Tripathi2020b}
S.~Tripathi, H.~Levine, M.~K. Jolly, {The Physics of Cellular Decision Making
  During Epithelial–Mesenchymal Transition}.
\newblock {\it Annu. Rev. Biophys.\/} {\bf 49}, 1--18 (2020).

\bibitem{Wells2015}
D.~K. Wells, W.~L. Kath, A.~E. Motter, {Control of Stochastic and Induced
  Switching in Biophysical Networks}.
\newblock {\it Phys. Rev. X\/} {\bf 5}, 031036 (2015).

\bibitem{Zanudo2017}
J.~G.~T. Za{\~{n}}udo, G.~Yang, R.~Albert, {Structure-based control of complex
  networks with nonlinear dynamics}.
\newblock {\it Proc. Natl. Acad. Sci. USA\/} {\bf 114}, 7234--7239 (2017).

\bibitem{Sullivan2023}
E.~Sullivan, M.~Harris, A.~Bhatnagar, E.~Guberman, I.~Zonfa, E.~{Ravasz Regan},
  {Boolean modeling of mechanosensitive epithelial to mesenchymal transition
  and its reversal}.
\newblock {\it iScience\/} {\bf 26}, 106321 (2023).

\bibitem{Hornos2005}
J.~E.~M. Hornos, D.~Schultz, G.~C.~P. Innocentini, J.~Wang, A.~M. Walczak,
  J.~N. Onuchic, P.~G. Wolynes, Self-regulating gene: An exact solution.
\newblock {\it Phys. Rev. E\/} {\bf 72}, 051907 (2005).

\bibitem{Walczak2005}
A.~M. Walczak, J.~N. Onuchic, P.~G. Wolynes, Absolute rate theories of
  epigenetic stability.
\newblock {\it Proc. Natl. Acad. Sci. USA\/} {\bf 102}, 18926--18931 (2005).

\bibitem{Tkacik2012}
G.~Tka{\v{c}}ik, A.~M. Walczak, W.~Bialek, {Optimizing information flow in
  small genetic networks. III. A self-interacting gene}.
\newblock {\it Phys. Rev. E\/} {\bf 85}, 041903 (2012).

\bibitem{Assaf2013}
M.~Assaf, E.~Roberts, Z.~Luthey-Schulten, N.~Goldenfeld, {Extrinsic Noise
  Driven Phenotype Switching in a Self-Regulating Gene}.
\newblock {\it Phys. Rev. Lett.\/} {\bf 111}, 058102 (2013).

\bibitem{Zhang2013}
K.~Zhang, M.~Sasai, J.~Wang, Eddy current and coupled landscapes for
  nonadiabatic and nonequilibrium complex system dynamics.
\newblock {\it Proc. Natl. Acad. Sci. USA\/} {\bf 110}, 14930--14935 (2013).

\bibitem{Bhattacharyya2020}
B.~Bhattacharyya, J.~Wang, M.~Sasai, Stochastic epigenetic dynamics of gene
  switching.
\newblock {\it Phys. Rev. E\/} {\bf 102}, 042408 (2020).

\bibitem{mussel2010boolnet}
C.~M{\"u}ssel, M.~Hopfensitz, H.~A. Kestler, Bool{N}et—an {R} package for
  generation, reconstruction and analysis of {B}oolean networks.
\newblock {\it Bioinformatics\/} {\bf 26}, 1378--1380 (2010).

\bibitem{Barrett2013}
T.~Barrett, S.~E. Wilhite, P.~Ledoux, C.~Evangelista, I.~F. Kim,
  M.~Tomashevsky, K.~A. Marshall, K.~H. Phillippy, P.~M. Sherman, M.~Holko,
  A.~Yefanov, H.~Lee, N.~Zhang, C.~L. Robertson, N.~Serova, S.~Davis,
  A.~Soboleva, {NCBI} {GEO}: {Archive} for functional genomics data
  sets—update.
\newblock {\it Nucleic Acids Res.\/} {\bf 41}, D991--D995 (2013).

\bibitem{Efron1985}
B.~Efron, R.~Tibshirani, {The Bootstrap Method for Assessing Statistical
  Accuracy}.
\newblock {\it Behaviormetrika\/} {\bf 12}, 1--35 (1985).

\bibitem{Mori2017}
F.~Mori, A.~Mochizuki, {Expected Number of Fixed Points in Boolean Networks
  with Arbitrary Topology}.
\newblock {\it Phys. Rev. Lett.\/} {\bf 119}, 028301 (2017).

\bibitem{Rozum2021}
J.~C. Rozum, J.~G{\'o}mez Tejeda~Za{\~n}udo, X.~Gan, D.~Deritei, R.~Albert,
  Parity and time reversal elucidate both decision-making in empirical models
  and attractor scaling in critical {Boolean} networks.
\newblock {\it Sci. Adv.\/} {\bf 7}, abf8124 (2021).

\bibitem{Gan2018}
X.~Gan, R.~Albert, {General method to find the attractors of discrete dynamic
  models of biological systems}.
\newblock {\it Phys. Rev. E\/} {\bf 97}, 042308 (2018).

\bibitem{Justice2014}
S.~S. Justice, A.~Harrison, B.~Becknell, K.~M. Mason, {Bacterial
  differentiation, development, and disease: Mechanisms for survival}.
\newblock {\it FEMS Microbiol. Lett.\/} {\bf 360}, 1--8 (2014).

\bibitem{Smits2006}
W.~K. Smits, O.~P. Kuipers, J.-W. Veening, {Phenotypic variation in bacteria:
  The role of feedback regulation}.
\newblock {\it Nat. Rev. Microbiol.\/} {\bf 4}, 259--271 (2006).

\bibitem{mcclure2013computational}
R.~McClure, D.~Balasubramanian, Y.~Sun, M.~Bobrovskyy, P.~Sumby, C.~A. Genco,
  C.~K. Vanderpool, B.~Tjaden, {Computational analysis of bacterial RNA-Seq
  data}.
\newblock {\it Nucleic Acids Res.\/} {\bf 41}, e140 (2013).

\end{thebibliography}
\bibliographystyle{ScienceAdvances}

\noindent \textbf{Acknowledgements:} \\

\noindent \textbf{Funding:} The authors acknowledge support from NSF grant MCB-2206974 and the use of Quest high performance computing facility at Northwestern University.\\
\noindent \textbf{Author Contributions:} TPW and AEM developed the research concept. YZ implemented the core algorithm, conducted the numerical simulations, and analyzed the results with assistance from TPW. TPW gathered and analyzed the sequencing data. KAR provided interpretation of the biological relevance. TPW and AEM led the writing of the manuscript. All authors reviewed the manuscript and approved the final version.\\
\noindent \textbf{Competing Interests:} The authors declare that no competing interests exist. \\
\noindent \textbf{Data and materials availability:} All data necessary for the reproduction of the reported results are included in the \nameref{sec:SM}. 
Software for employing the method is available from Zenodo (DOI: \url{10.5281/zenodo.12775479}) and additionally on GitHub (\url{https://github.com/yizhao-nu/Irreversiblility-in-GRN}). Example results for generating the figures are available on Dryad (DOI: \url{https://doi.org/10.5061/dryad.547d7wmhg}).

\clearpage

\clearpage
\newpage
\setcounter{page}{1}

\refstepcounter{section}
\setcounter{section}{0}
\refstepcounter{figure}
\setcounter{figure}{0}
\renewcommand{\thefigure}{S\arabic{figure}}
\refstepcounter{table}
\setcounter{table}{0}
\renewcommand{\thetable}{S\arabic{table}}
\setcurrentname{Supplementary Material}
\label{sec:SM}

\begin{center}

\section*{\fontsize{17pt}{1.1}\selectfont Supplementary Material: Irreversibility in Bacterial Regulatory Networks}

{\large Yi Zhao, Thomas P. Wytock, Kimberly A. Reynolds, and Adilson E. Motter}\\
Corresponding Author: Adilson E. Motter.
E-mail: motter@northwestern.edu
\end{center}
\bigskip
\clearpage
\newpage
\subsection*{Summary}
\noindent This Supplementary Material file contains details regarding 
(i) the consistency of the irreversibility estimates across realizations, 
(ii) a characterization of the nodes not exhibiting irreversibility, 
(iii) a discussion of the necessity of positive circuits for irreversibility, 
(iv) conditions necessary for the preservation of synchronous partial fixed points under asynchronous updates,
(v) an examination of the network motifs responsible for certain periodic attractors,
(vi) a comparison with alternative formulations of the ensemble of rules,  
(vii) a comparison with previous studies investigating bacterial heterogeneity, 
(viii) a comparison of adaptive evolution responses to \textit{crp} KO with irreversible response genes in the Boolean model, and
(ix) a detailed description of how to use our results to obtain specific experimental predictions.

\subsection*{Irreversibility across realizations of the update rules}
We demonstrate that the averages for the irreversibility estimated from different samplings of the rules converge to a common value. 
\Cref{fig:rules_ensemble} presents the distributions of the RMSD for all considered parameter pairs with nonunique rules for ascending and descending input sortings (i.e., diffuse and concentrated control) in panels A and B, respectively.  
The plots in each row are arranged in order of decreasing rule bias [cf., Fig.~\ref{fig:rule-contours}A]. Importantly, neither the rule bias nor the average canalization depth appears to alter the convergence. 
We observe that the RMSD 
converges to approximately $0.1$ in probability for two independent ensembles of $M'=10$ realizations, with slightly higher values observed in ascending order compared to descending order. 
These results establish that $M=20$ realizations are sufficient to obtain reliable estimates of the average irreversibility. In the main text, the ensemble averages over realizations are compared with irreversibility for unique rules, which do not require averaging.

In addition to considering the impact of update rules, we also examined the impact of changing the point of the attractor in which the perturbation was reverted in the case of periodic attractors.
Throughout the paper, we assess irreversibility by applying the perturbation to the first point of each attractor recorded by the SAT algorithm~\cite{dubrova2011sat}. Similarly, we revert the perturbation in the first point of the perturbed trajectory that is on the new attractor. 
We argue that the impact of this choice is small, which we confirm
by comparing with the irreversibility observed when reverting the perturbation of the closest point of the new attractor to the initial state in terms of Hamming distance. We find that reverting from this state causes 2.1\% of irreversible transitions to become reversible.

\subsection*{Nodes not exhibiting irreversibility}
\label{sec:rev}

The algorithm that generates the core network $G'$ recursively trims nodes that are trivial SCCs, so that the core network contains all irreversible perturbations.
This is the case because 
if all inputs of an SCC consisting of a single node $v$ with no autoregulation remain unchanged after a transient perturbation, then $x_v$ must remain unchanged as well.
Thus, it can be concluded that the trimming will not alter the number of fixed points in the network,
which is consistent with known necessary conditions for multi-stationarity~\cite{Angeli2004,Remy2008,Craciun2011,Mori2017}.
Now focusing on the core, Fig.~\ref{fig:result2} shows that  $36$ nodes within $G'$ do not admit irreversible perturbations across all realizations of the rules in our simulations. 
We examine these nodes on a case-by-case basis and show that they cannot be irreversible by directly demonstrating that the final state must belong to the same attractor as the initial state. 
There are two cases: 

\begin{enumerate}
    \item \textit{Reversibility of leaf nodes.} Consider a leaf node $u$, which by definition has no edges to other nodes. Such a node is necessarily reversible because it does not influence other nodes and is restored to its original state upon removal of the transient perturbation. A total of 32 nodes are in this class---\textit{aidB}, \textit{araC}, \textit{asnC}, \textit{betI}, \textit{cadC}, \textit{cusR}, \textit{dnaA}, \textit{dpiA}, \textit{fucR}, \textit{glcC}, \textit{glnG}, \textit{hyfR}, \textit{idnR}, \textit{lldR},  \textit{lsrR}, \textit{malI}, \textit{melR}, \textit{metR}, \textit{mraZ}, \textit{nhaR}, \textit{nikR}, \textit{pdeL}, \textit{prpR}, \textit{purR}, \textit{putA}, \textit{puuR}, \textit{rbsR}, \textit{tdcA}, \textit{yeiL}, \textit{yiaJ}, \textit{yqjI}, and \textit{zraR}.
    
    \item \textit{Reversibility of nodes exclusively regulating autorepressive leaf nodes.} The genes \textit{metJ}, \textit{pdhR}, \textit{argP}, and \textit{nac} have a single regulatory output to an autorepressive leaf node. 
    Either the regulatory node canalizes the output of the leaf node or the leaf node's autoregulatory input (which is by definition 0) overrides the other input. 
    In the former case, any change to the state of the leaf node caused by the perturbation of the regulatory node is restored by its reversion, whereas in the latter case the state of the leaf node never changes.
\end{enumerate}

\subsection*{Necessity of positive circuits for irreversibility}
\label{sec:nec_pos}
In the main text, we focus on positive circuits and claim that these
circuits underlie the multistability necessary for irreversibility. 
Here we justify this claim.
\textit{We first observe that the edge consistency condition implies the existence of at least one circuit in the network.} 
The edge consistency condition requires that each node in the network has an incident edge (this assumes the existence of autoregulatory edges in nodes without input edges from other nodes).
It follows that in a connected component of $|V|$ nodes, starting at an arbitrary node and following incident edges backwards $|V|$ times, we necessarily visit a node more than once. 
This implies the existence of a circuit.

\textit{We next observe that circuits are necessary for multistability and, by implication, for irreversibility}. This is because nodes not belonging to circuits are trivial SCCs and thus are not irreversible response nodes according to the argument in the previous section. 
The \textit{phoB} origon core network has numerous positive circuits, guaranteeing multiple stable attractors.

Irreversibility occurs between two attractors.
To orient the discussion, we note that in our simulations, 42.2\% of the attractors are period-1 attractors, 24.4\% are period-2 attractors, 1.1\% are period-3 attractors, 27.9\% are period-4 attractors, and 4.4\% are higher-period attractors. 
For brevity, we use the language \textit{periodic attractors}
to refer to period-$n$ attractors with $n\!>\!1$, while we continue to use \textit{fixed points} to refer to period-1 attractors.
Irreversibility may involve a transition between two fixed-point attractors, two periodic attractors, or a fixed-point attractor and a periodic attractor.

We first consider the case of irreversibility arising from a transition between two fixed-point attractors, noting that 49.0\% of transitions in our simulations are of this type.
\textit{We observe that fixed-point attractors cannot arise from negative circuits}. In a network consisting of a negative circuit, the fixed-point condition $\mathbf{x} = \mathbf{B}(\mathbf{x})$ implies that the state $x_v$ would be negated an odd number of times in the process of applying the update rules \cite{Remy2008}, which would require that $x_v = \bar{x}_v$ (we recall that the overbar indicates negation). 
Since this is impossible, no fixed-point attractors can exist for a network consisting of a negative circuit. 
In the absence of positive circuits, application of the fixed-point condition to a network including any number of negative circuits will yield one or more contradictions of the form $x_v = \bar{x}_v$, which again exclude the possibility of fixed-point attractors. 
In the absence of positive circuits, the exceptions are: (i) negative loops consisting of a single node with an autorepressive edge, which is fixed but monostable, 
and (ii) negative circuits that remain fixed due to inputs from upstream nodes, which precludes the possibility of multistability.
Thus, irreversibility between fixed points requires positive circuits.

We note that periodic attractors in which \textit{all} nodes are time-dependent are absent in our simulations. 
Thus, transitions in which one or more attractors are periodic involve \textit{partial} fixed points, which we define as time-dependent attractors in which the state of one or more nodes are time-independent.
A total of 31.7\% of irreversibility transitions in our simulations occur between partial fixed-point attractors with all differences taking place among the time-independent nodes. 
An additional 2.2\% involve a transition between a fixed-point and a partial fixed-point attractor.
The remaining transitions (17.1\%) occur between partial fixed-point attractors with different numbers of fixed nodes, meaning that a subset of the time-independent nodes transition to being time-dependent and/or vice versa.
In all cases, the differences between attractors involve a node that is time-independent in at least one of the attractors. 
We observe that the arguments of the previous paragraph apply in the sense that positive circuits determine the time-independent nodes in partial fixed-point attractors.
Thus in all transitions, the attractors are determined (at least in part) by positive circuits.

\textit{We expect the time-independent portions of the initial and final attractors observed in our simulations to be preserved under asynchronous updates in the large majority of cases.}
This is the case because fixed-point attractors are preserved under asynchronous updates~\cite{Rozum2021}, implying that the initial and final states in the 49.0\% of transitions between fixed-point attractors will exist in asynchronous updates as well.
In addition, we expect the time-independent nodes of partial fixed points to be preserved in a further 16.2\% of transitions, 
because a synchronous partial fixed point has a corresponding asynchronous partial fixed point under the conditions discussed below.
Finally, the initial and final attractors in transitions involving one fixed point and one partial fixed point are preserved, and they are guaranteed to remain distinct 
when the partial fixed point has time-independent nodes whose state differs from those of the fixed point. The latter accounts for 2.6\% of all transitions. 
Together, 67.9\% of transitions for synchronous updates have the initial and final attractors preserved under asynchronous updates.

\textit{The time-dependent portions of partial fixed-point attractors associated with negative circuits show acyclic behavior under asynchronous updates.}
This is because the periodicity of an attractor generally requires the concurrent update of multiple nodes at each time step, while only one (randomly chosen) node is updated at a time under asynchronous updates. 
However, we did not observe transitions between attractors 
in which the only difference occurs between time-dependent nodes
(i.e., attractors that share the same set of time-independent nodes and that have identical states on these nodes) in our simulations.

\subsection*{Preservation of synchronous partial fixed points}
Asynchronous partial fixed points imply the existence of synchronous partial fixed points~\cite{Gan2018} but not vice versa. 
Concerning the forward implication, the states of time-independent nodes of asynchronous partial fixed points are, by definition, guaranteed to remain fixed for all possible states of the time-dependent nodes, and thus under synchronous updates.
The reverse implication does not hold
because the asynchronous updates cause the time-dependent nodes to reach states outside of their synchronous periodic orbit, making it possible for the downstream time-independent nodes to change.

Notwithstanding, we identify a set of sufficient conditions under which synchronous partial fixed points have corresponding asynchronous partial fixed points. 
For the state of a time-independent node in a synchronous partial fixed point to change under asynchronous updates, the node must have more than one upstream input that is time-dependent. Otherwise, the time evolution on the attractor is sufficient to conclude that the time-independent state will be stable for all possible values of the time-dependent nodes. 
The requirement for multiple time-dependent inputs is established by as follows.
We refer to the set of time-dependent nodes in the attractor as $\mathcal{B}^{0}$ and the set of nodes incident on $u$ in $G'$ as $\mathcal{I}_u$. 
Then, the set of nodes with at least 2 incident nodes in $ \mathcal{B}^{0}$ is $d\mathcal{B}^{0} = \{u\,\big|\, 1 < |\mathcal{I}_u \cap \mathcal{B}^0| \}$.
Starting at $i=0$, we set $\mathcal{B}^{i+1}$ to $\mathcal{B}^{i} \cup d\mathcal{B}^{i}$, increment $i$ by 1, and iterate the calculation until $d\mathcal{B}^{i}$ is empty.
Thus, the time-independent nodes not in $\mathcal{B}^{i}$ are preserved under asynchronous updates. 
As a corollary, this implies that $|d\mathcal{B}^{0}|=0$ is a sufficient condition to conclude that synchronous partial fixed points imply the existence of corresponding synchronous partial fixed points.
This condition is met in 82.5\% of the period-2 attractors,
88.9\% of the period-3 attractors,
14.4\% of the period-4 attractors,
and 37.6\% of the higher-period attractors
of our simulations. 
Together, the time-independent nodes of partial fixed-point attractors are guaranteed to be preserved under asynchronous updates in at least 46.3\% of cases. (This is calculated by multiplying the frequency of each periodic attractor times its rate of preservation and normalizing by the total number of periodic attractors). 
If the initial and final attractors in irreversible transitions were uncorrelated, then 21.4\% of transitions between partial fixed-point attractors would be preserved under asynchronous updates. 
We find that 51.2\% of transitions between partial fixed points have initial and final attractors that preserved in our simulations. 
This percentage applies to the 31.7\% of transitions between partial fixed points for which both the initial and final states have the same set of time-dependent nodes, yielding a total of 16.2\% of all transitions as referenced above.

\subsection*{Motifs generating periodic attractors}
The analysis of partial fixed points motivates us to investigate the network structures behind the period-2 and period-4 attractors, which together account for over half of the attractors in our simulations.
We attribute the prevalence of period-2 and period-4 attractors specific motifs in the network.
Period-2 attractors occur mainly among gene pairs that exhibit mutual negative feedback such as
(\textit{galR}, \textit{galS}), (\textit{mlc}, \textit{ptsG}), (\textit{exuR}, \textit{uxuR}), and (\textit{mazE}, \textit{mazF}).
Period-4 attractors occur among gene pairs in which
the first represses the second and the second activates the first, and one of the genes has no autoregulation, as in the cases of
(\textit{arcA}, \textit{fnr}) and (\textit{hns}, \textit{cspA}). 
These pairs can have a period-4 orbit $(1,0) \rightarrow (0,0) \rightarrow (0,1) \rightarrow (1,1) \rightarrow (1,0)$ that cascades to downstream nodes.

\subsection*{Comparison with alternative formulations}
Here, we offer evidence that the conclusions of the paper apply to the transcriptional regulatory network of \emph{E. coli} in general, even though we focus specifically on the \emph{phoB} origon (the largest origon) and canalizing rules. Specifically, we examine how changing the origon, the rules, and/or the mathematical formulation of the dynamics would affect the results.

\paragraph*{Different origons.}
In Table~\ref{tab:origon_overlap}, we report the statistics of all origons with core size $\geq 30$ in our RegulonDB model (the remaining origons have core sizes $\leq 5$). 
For our purposes, the most relevant measures are the overlaps of the core \emph{phoB} origon with those of the alternative origons, 
as well as the overlaps of nodes in the nontrivial SCCs of the \emph{phoB} origon with those of the alternative origons. 
As demonstrated in the main text, the ability of a node to admit an irreversible perturbation is related to its proximity to SCCs with positive circuits (in the general case, nontrivial SCCs can be made up of positive and/or negative circuits). 
Our model shows that most such SCCs of the largest origons are contained within the  \textit{phoB} origon core network (see overlap column in Table~\ref{tab:origon_overlap}). 
Accordingly, the identity and irreversibility probability of nodes that admit irreversible perturbations  
in our analysis of the \textit{phoB} origon core network are inclusive of almost all cases that would be found when considering all largest origons.

\paragraph*{Threshold-based rules.}

One alternative to canalizing dynamics is threshold-based dynamics~\cite{Tripathi2020}, in which the update rules for each node are
\begin{equation}
    x^{t+1}_u = \begin{cases}
        1 \ \mathrm{if}\ \sum_{i =1}^{k^+_u} y^t_{v_i} / k^+_u > \chi_u, \\
        0 \ \mathrm{otherwise},
    \end{cases} \label{eq:threshold}
\end{equation}
where $\chi_u \in [0,1]$ are activation thresholds. 
Specializing to the case of a uniform threshold $\chi_u=\chi$, the limit  $\chi \rightarrow 0$ corresponds to the  case $(r,s)=(0,1)$ (i.e., all inputs joined by OR operators) whereas the limit
$a \rightarrow 1$  corresponds to $(r,s)=(0,0)$ (i.e., all inputs joined by AND operators).
The sorting of inputs $y_{v_i}$ by $k^-_{v_i}$ when determining canalization also has
a relationship to threshold dynamics with heterogeneous $a_u$. In particular, if $\chi_{v_i} \propto k_{v_i}^-$,
then nodes with small $k_{v_i}^-$ will be
easier to activate. The ease of activation of these nodes makes this
case analogous to the ascending input sorting in the limit that  $s \rightarrow 1$ when $r=0$. 
Other heterogeneous schemes with  $\chi_u$ drawn uniformly from $[0,1]$ would correspond most closely to the cases of $(r,s)=(0.4,1)$ and $(r,s)=(0.6,1)$ in our simulations.

\paragraph*{Bias-based rules.} 
Bias-based rules are a formulation of random Boolean networks in which 
the update rules are generated by randomly assigning $1$ (with probability $\nu$) or $0$ to each possible input vector.
The parameter $\nu$ is equivalent to the rule bias used in our analysis.
This formulation is mathematically convenient for examining the effect of topology on the stability of Boolean networks~\cite{Pomerance2009,Squires2014}. 
Bias-based rules apply in the case of tree-like graphs in which the activation probabilities of nodes may be regarded as independent, and they do not rely on information about the polarity of interactions. 
In this work, we operate in the opposite limit: we focus specifically on the core network, where the coregulation of nodes produces correlated activations, and
our empirical network allows us to explicitly incorporate edge polarity into the model. 

\paragraph*{Alternative dynamical formulations.}
We argue that the predictions of the model apply beyond the case of Boolean rules that are synchronously updated. 
Related conclusions are expected for asynchronous updates 
since fixed points are known to be preserved~\cite{Rozum2021} and, as shown above, most partial fixed-points are also preserved under asynchronous update rules.
Similar conclusions apply to 
continuous representations of the dynamics because the rules can be transformed from a discrete to a continuous representation of the state space and time that preserves the stable steady states~\cite{Wittmann2009}. We examine the outcomes of such a transformation for a simple system below in an effort to develop experimental predictions that are as specific as possible.
Given the correspondence between our results and continuous systems, it is natural to conjecture that transient perturbations may lead to 
irreversibility in stochastic models of gene regulation that account for the transcription, translation, and degradation~\cite{Hornos2005,Walczak2005,Tkacik2012,Assaf2013,Zhang2013,Bhattacharyya2020}. 
This conjecture may be investigated by mapping the irreversible perturbations in our simulations into corresponding parameter changes in the stochastic models.
 
\subsection*{Comparing with bacterial differentiation and environmental perturbation} 
It is instructive to compare the irreversibility from \textit{transient genetic perturbations} with bacterial differentiation processes, such as sporulation in \textit{B. subtilus}~\cite{Veening2008},  stalking in \textit{C. crescentus}~\cite{Shapiro1971}, and life-cycle stages in pathogenic \textit{E. coli}~\cite{Justice2014}.
The results can also be compared with observations of
bacterial heterogeneity such as 
persistence~\cite{Smits2006} 
across multiple bacterial species,
the mucoid phenotype of \textit{P. aeruginosa}~\cite{Smits2006}, 
competence in \textit{B. subtilus}~\cite{Veening2008}, 
and 
metabolic shifts in \textit{E. coli} in the case of the \textit{lac} operon~\cite{Ozbudak2004} and other inducible sugars~\cite{Afroz2014}.
In these examples, genetically identical strains can exhibit diverse phenotypes due to \textit{environmental cues},
and these phenotypes can persist hysteretically after the environmental cues are removed. 
While these examples involve environmental changes, our analysis shows that regulatory switching alone is sufficient to alter the state of positive feedback loops, resulting in irreversibility in the transcriptional state of the cell.
Conversely, by relating our analysis with these examples, it follows that the transcriptional irreversibility we characterize can give rise to large phenotypic changes, including morphological and behavioral ones. 
This transcriptional irreversibility may be leveraged to design synthetic bacterial circuits that generate irreversible changes, emulating recent developments in 
mammals~\cite{Zhu2022}.

\subsection*{Comparison with adaptive evolution responses}

In the main text, we explore the question of 
how similar the set of irreversible response genes to the transient perturbation of \textit{crp} KO is to  the set of genes exhibiting altered expression in the adaptive evolution response to the same KO. 
In \cref{tab:batch_adapt,tab:chemo_adapt}, we list the identity of the genes with the largest shifts in expression ($\ln \rho_u / \langle \rho \rangle$) when the adapted strain is cultivated in batch and chemostat conditions, respectively.
To facilitate a comparison between our model and the data, we include the sign of regulation of \textit{crp} ($\sigma_u^{\rm mod}$), which is supposed to have the \textit{opposite} sign as the shift in expression. As noted in the main text, 10 of 11 genes in batch cultivation and 33 of 42 genes in chemostat cultivation have changes in expression consistent with the regulatory model.
The majority of these genes are found to respond irreversibly in our model. 
Of the genes that do not, most are directly regulated by \textit{crp} and negatively autoregulated (\textit{malI}, \textit{lsrR}, \textit{glcC}, \textit{prpR}, \textit{rbsR}, and \textit{glnG}), 
two are negatively autoregulated but indirectly regulated by \textit{crp} (\textit{metR} and \textit{purR}), and
two are not regulated by \textit{crp} (\textit{pdeL} and \textit{cra}). 
Changes in the first two cases may be due to constitutive activity of the gene promoters, which is not accounted for by our model.
The final case may reflect changes to the intracellular environment not captured by the model.
The sign of the gene expression responses to adaptive evolution are significantly concordant with those predicted by the model (see main text). 
In addition, the set of genes is enriched for irreversible response genes to \textit{crp}.

\subsection*{Strategy to identify specific experimental conditions}

\begin{figure}[h!]\small\centering
\includegraphics[width=0.67\textwidth]{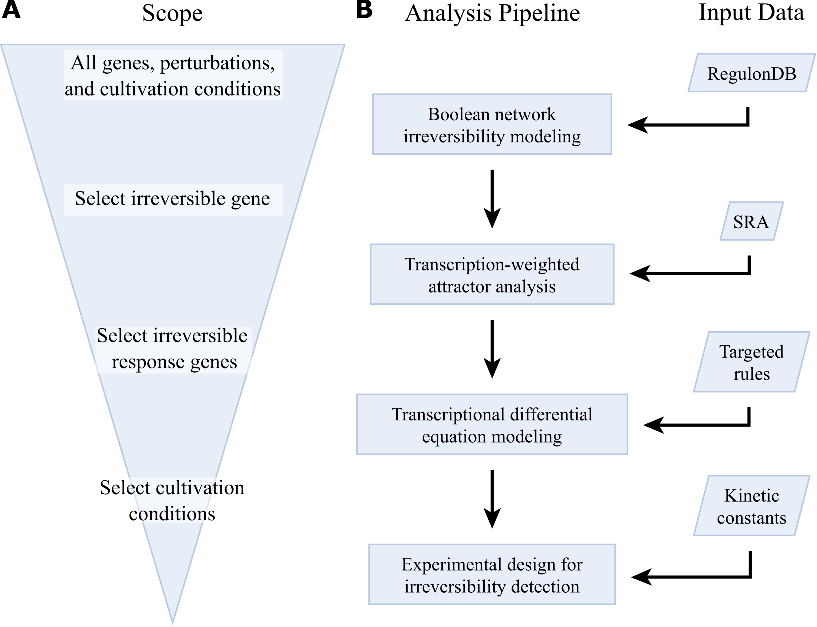}
\caption{  \textbf{Schematic for designing experiments to find irreversibility.} (\textbf{A}) Diagram of how the scope is narrowed from all possible gene perturbations, irreversible response genes, and cultivation conditions to a manageable number of experimental predictions. (\textbf{B}) Analysis and input data needed to narrow the scope at each stage. 
}
\label{fig:scope_schematic}
\end{figure}

Our Boolean framework, presented in the main text, fits into a broader strategy to identify specific conditions under which irreversibility occurs.
An experiment to observe irreversibility requires the specification of (i) a candidate irreversible gene perturbation, (ii) an associated irreversible response gene, and (iii) specific cultivation conditions.
\Cref{fig:scope_schematic}A summarizes how the scope of potential experiments narrows. 
Then, the flow chart in \cref{fig:scope_schematic}B explains how the various levels of modeling (which themselves are based on different kinds of input data) narrow the scope.
With (i)--(iii) specified, the designed experiment would employ CRISPR-interference to carry out the gene perturbation and quantitative polymerase chain reaction (qPCR) or RNA-seq to measure
the expression irreversible response gene at different time points.

\subsubsection*{Overview}
One major outcome of the Boolean modeling is the prominence of \textit{crp} KO as a perturbation that leads to irreversibility. (The fact that targeted KOs have fewer externalities on the use of cellular resources than OEs encouraged us to focus specifically on transient \textit{crp} KO.)
This motivated us to examine the Sequencing Read Archive (SRA) for \textit{E. coli} experiments in which \textit{crp} is perturbed. 
We found 16 instances of experiments characterizing \textit{crp} downregulation either as a result of genetic perturbations or environmental shifts.
Given this information, we examined the attractors that admit irreversibility of \textit{crp} KO to determine whether they were similar to experimentally observed states, as described in the subsection on transcription-weighted attractor analysis.
Briefly, predictions of irreversible response genes associated with attractors that were more similar to experimentally observed states were given higher weights than those that were more divergent. 
The outcome of the this analysis was a set of autoexcitatory genes that are themselves positively regulated by \textit{crp}. 
This is similar to the three-gene schematic in \cref{fig:nec-suff}, except that the two-gene positive feedback loop is reduced to a single autoexcitatory gene. 
Note that from the Boolean modeling, we already know that an AND operator is required for irreversibility (i.e., expression of both genes is required for activation).
This requirement specifies the form of the differential equation model needed to model the candidate irreversible genes, as discussed in the subsection on transcriptional differential equation modeling.
From these differential equations, it is possible to identify relative values of the parameters necessary for irreversibility to occur and thereby specify the cultivation conditions needed to observed irreversibility.

\subsubsection*{Transcription-weighted attractor analysis}
\paragraph*{Methods:} We searched the GEO database~\cite{Barrett2013} for all ``Expression profiling by high-throughput sequencing'' assays conducted in ``\textit{Escherichia coli}'' on November 15, 2022, which yielded 495 datasets. 
Of these, we retained datasets that (i) had at least 10 RNA-seq samples and (ii) were conducted on non-enterohemorrhagic \textit{E. coli}, yielding 155 datasets with 5,147 gene expression profiles.  
We retrieved the raw RNA-seq data from SRA, aligned it to the MG1655 genome (NCBI Reference: NC\_000913.3) using Rockhopper~\cite{mcclure2013computational}, and calculated the TPM. The TPM calculations excluded alignments to ribosomal RNA genes.

We next extracted the TPM counts for the 87 genes in the core network and converted them to log-scale using the formula $\zeta_i = \log_{10}(z_i + 10^{-10})+10$, where $i$ is an index over genes. We subsequently identified 16 instances in which \textit{crp} downregulation was measured within a dataset. 
For each case, we calculated the average expression of each gene before ($t = O$) and after ($t = Q$) \textit{crp} downregulation. 
The average expression was binarized to 1 if it was larger than the gene's median in the entire dataset and 0 otherwise.
Next, we compared the observed expression states before and after \textit{crp} downregulation with the attractors generated in our Boolean model before and after \textit{crp} KO. 
As in the Boolean modeling, we use $\mathbf{x}$ to represent the binarized expression obtained from data.
Furthermore, we use $x_i^{\rm obs}$ and $x_i^{\rm att}$ to denote the attractors obtained from experimental observations and our Boolean modeling, respectively, and we define $\phi_i$ to be the fraction of cases for which $x_i=1$ in the expression data. 
Then, the similarity between observed and modeled states is quantified using the Hamming similarity 
\begin{equation}
  D(\mathbf{x}^{\rm obs}, \mathbf{x}^{\rm att}) = \exp\left(-\sum_{i=1}^{|V'|} | x_i^{\rm obs} - x_i^{\rm att}|\right)   \label{eq:Hamming}
\end{equation}
and likelihood of $x_i^{\rm obs}$ matching $x_i^{\rm att}$ if each $x_i$ is an independently assigned Bernoulli random variable with parameter $\phi_i$:
\begin{align}
    E(\mathbf{x}^{\rm obs}, \mathbf{x}^{\rm att}, \bm{\phi}) &= \prod_{i=1}^{|V'|} \theta_i(x_i^{\rm obs},\phi_i)^{\kappa_i(x_i^{\rm obs}, x_i^{\rm att})}, \ \quad\quad{\rm where}\label{eq:LOM} \\
    \theta_i(x_i^{\rm obs},\phi_i)&=\begin{cases}
        \phi_i\quad\quad &{\rm if}\ x_i^{\rm obs} = 1,\ {\rm and} \nonumber \\
        1-\phi_i\ &{\rm if}\ x_i^{\rm obs} = 0, \nonumber
    \end{cases}\ \quad{\rm with}  \\
    \ \kappa_i(x_i^{\rm obs}, x_i^{\rm att}) &= \begin{cases}
        1 &\quad\ {\rm if}\ x_i^{\rm obs}=x_i^{\rm att},\ {\rm and} \nonumber \\
        -1 &\quad\ {\rm if}\ x_i^{\rm obs}\neq x_i^{\rm att}.
    \end{cases}
\end{align}
The irreversible responses are averaged using \cref{eq:Hamming,eq:LOM} according to the following procedure: 
\begin{enumerate}
    \item[(i)] Identify the set of all attractors across all rule realizations in which $x_{crp}=1$, denoted $\mathcal{C}^{\rm on}$,  and indexed by $j \in \{1,...,|\mathcal{C}^{\rm on}|\}$.
    \item[(ii)] Characterize the genes' responses to \textit{crp} KO in binary matrices $\Delta \in \{0,1\}^{|\mathcal{C}^{\rm on}| \times |V'|}$ and $\Psi \in \{0,1\}^{|\mathcal{C}^{\rm on}| \times |V'|}$ where $\Delta_{j\ell} = 1$ if the $\ell$th gene responds irreversibly in attractor $j$ and $\Psi_{j\ell} = 1$ if the $\ell$th gene responds reversibly in attractor $j$.
    \item[(iii)] For each
instance of \textit{crp} downregulation, indexed by $k \in \{1,...,16\}$, fix $\mathbf{x}^{(k,O)}$ and $\mathbf{x}^{(k,Q)}$ and calculate the attractor weights $A_{jk} = \prod_{t \in \{O,Q\}} D(\mathbf{x}^{(k, t)},\mathbf{x}^{(j, t)})$ or $A_{jk} = \prod_{t \in \{O,Q\}} E(\mathbf{x}^{(k, t)},\mathbf{x}^{(j, t)},\bm{\phi})$.
    \item[(iv)] With $k$ still fixed, normalize the $A_{jk}$ to $\tilde{A}_{jk} = A_{jk} / \sum_{j=1}^{|\mathcal{C}^{\rm on}|}A_{jk}$ and compute the average probability of irreversible and reversible response weighted by the $k$th conditions, which are given by $I_{k\ell} = \sum_{j=1}^{|\mathcal{C}^{\rm on}|} \tilde{A}_{jk} \Delta_{j\ell}$  and $R_{k\ell} = \sum_{j=1}^{|\mathcal{C}^{\rm on}|} \tilde{A}_{jk} \Psi_{j\ell}$, respectively.
    \item[(v)] Compute $\langle I_\ell \rangle = \sum_{k=1}^{16} I_{k\ell}$ and $\langle R_\ell \rangle = \sum_{k=1}^{16} R_{k\ell}$, the average of $I_{k\ell}$ and $R_{k\ell}$, respectively, across conditions.
\end{enumerate}
Using these steps, we calculated the weighted probability of responding reversibly and irreversibly for each gene.

\begin{figure}[h!]\small\centering
\includegraphics[width=\textwidth]{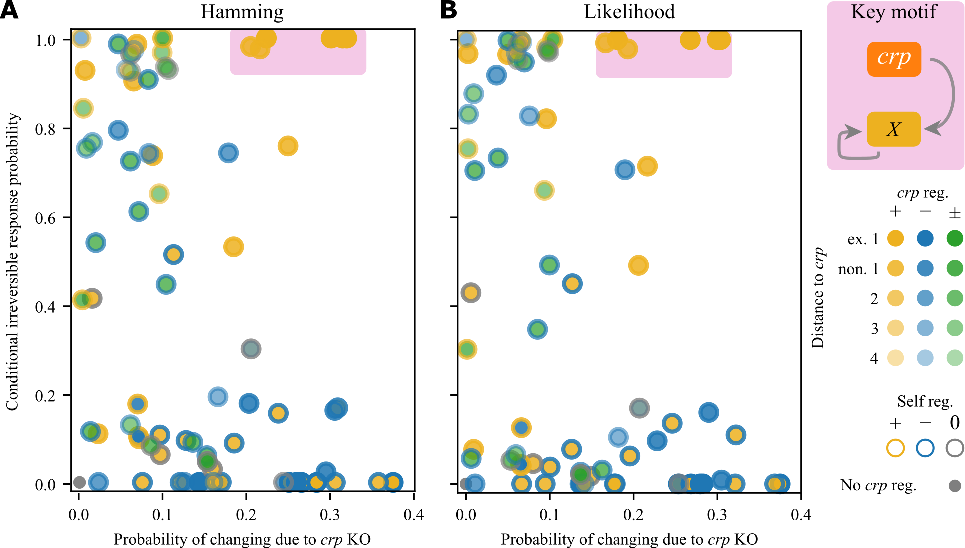}
\caption{ \textbf{Probability of irreversible response after weighting attractors by their similarity to observed states.} 
(\textbf{A}) Probability of responding irreversibly conditioned on a gene changing in response to \textit{crp} KO. Probabilities indicated on the vertical axis are expressed as a fraction of the corresponding values indicated on the horizontal axis. Attractors are weighted proportionally to the exponential of the negative Hamming distance (see \cref{eq:Hamming}). Node face colors correspond to the sign of \textit{crp} regulation, node edge colors correspond to the sign of self-regulation, and node opacities correspond to the distance from \textit{crp} as summarized in the legend at right.
(\textbf{B}) Same plot as in (A), when attractors are weighted by likelihood (see \cref{eq:LOM}).
In both panels, the genes in the pink box belong to the key motif in which \textit{crp} activates the gene \textit{X} and the gene \textit{X} activates itself. 
}
\label{fig:wt_attr_analysis}
\end{figure}

\paragraph*{Results:} \Cref{fig:wt_attr_analysis} summarizes the results of the probability for each gene to respond irreversibly when weighting the attractors by their similarity to observed transcriptional states. The horizontal axis indicates the fraction of \textit{crp} KO for which each gene changed state between the initial attractor and the attractor reached after \textit{crp} KO. Given that the gene changed, the vertical axis reports the fraction of instances that it remained altered in the final attractor. Comparing \cref{fig:wt_attr_analysis}A to \cref{fig:wt_attr_analysis}B reveals that the patterns of irreversibility as they relate to network structure are mostly preserved across the two weighting strategies. 
In particular, genes likely to respond irreversibly lie in the top right of each plot and are highlighted by a pink box. All such genes exhibit the key motif in which \textit{crp} activates the target gene (\textit{X}), which also activates itself. This result motivates an exploration of more granular differential equation models covered in the next section.

In the remaining sectors of \cref{fig:wt_attr_analysis}, we see the irreversibility as it relates to other regulatory architectures.  In the lower right, we see a group of autorepressive genes activated by \textit{crp} that turn off upon \textit{crp} KO, but are restored with \textit{crp} expression. Such genes would be candidates for negative controls in a high throughput experiment, since they are likely to respond, but not likely to be irreversible.
Moreover, as the distance of the candidate irreversible response genes to \textit{crp} increases, the probability of changing upon \textit{crp} KO decreases. 
This occurs because the response to the KO depends on a larger number of regulatory rules.
At the same time, once a change occurs, the probability that it is irreversible becomes less concentrated around 0 and 1.
This may be attributed to the larger number of ways to fit the candidate gene inside or downstream of a positive circuit for a given distance to \textit{crp}, regulatory sign, and self-regulation.

\subsubsection*{Transcription differential equation modeling}
\paragraph*{Analysis:} The key motif identified in \cref{fig:wt_attr_analysis} and the Boolean modeling together imply that irreversibility for the \textit{crp} KO requires that both \textit{crp} and \textit{X}  be present for \textit{X} to be activated (i.e., an AND operator). To transform this Boolean logic into a set of continuous differential equations, we use the multivariate polynomial interpolation method~\cite{Wittmann2009} and make the following approximations: (i) transcription is fast relative to translation, (ii) dilution dominates protein degradation as a mechanism for protein loss, and (iii) the basal transcription rate of gene $X$ is negligible. 
This yields the following system of equations for the expression of \textit{crp} ($C$) and the downstream gene ($X$):
\begin{align}
    \frac{dC}{dt} = & \alpha + \beta_C \frac{C^{\eta_C}}{K_{C}^{\eta_C} + C^{\eta_C}} - \delta C \label{eq:crp_unscaled} \\
    \frac{dX}{dt} = & \beta_X \left( \frac{C^{\eta_C}}{K_{C}^{\eta_C} + C^{\eta_C}}\right)\left( \frac{X^{\eta_X}}{K_{X}^{\eta_X} + X^{\eta_X}}\right) - \delta X \label{eq:X_unscaled},
\end{align}
where 
\begin{enumerate}
    \item $\alpha$ is the basal transcription rate of \textit{crp},
    \item $\beta_C$ and $\beta_X$ are transcriptional activation parameters (resulting from transcription factor binding),
    \item $\delta$ is the dilution rate,
    \item $\eta_C$ and $\eta_X$ are Hill coefficients (which designate more switch-like behavior as they become larger), and
    \item $K_C$ and $K_X$ are the concentrations for half-maximal activation.
\end{enumerate}
These phenomenological parameters are measurable by targeted experiments, and as such, they may guide the choice of gene $X$. Here, we assume for simplicity that the impact of the basal transcription of $X$ is negligible compared to the regulatory activation. This assumption does not impact the following analysis.

For the purposes of analyzing the parameters that lead to irreversibility, it is convenient to rescale the variables and parameters using
\begin{align}
    \bar{\alpha} = \frac{\alpha}{K_C} \quad \bar{\beta}_C =  \frac{\beta_C}{K_C} \quad \bar{\beta}_X = (1+\bar{C}^{-\eta_C}) \frac{\beta_X}{K_X} \quad
    \bar{C} = \frac{C}{K_C} \quad \bar{X} = \frac{X}{K_X} \label{eq:dim}.
\end{align}
Substitution of \cref{eq:dim} into \cref{eq:crp_unscaled,eq:X_unscaled} yields the following rescaled equations
\begin{align}
    \frac{d\bar{C}}{dt} = & f_C(\bar{C}) =  \bar{\alpha}_C + \bar{\beta}_C \left(1+\bar{C}^{-\eta_C}\right)^{-1} - \delta \bar{C} \label{eq:crp_scaled} \\
    \frac{d\bar{X}}{dt} = & f_X(\bar{X},\bar{C}) =  \bar{\alpha}_X + \bar{\beta}_X (1+\bar{X}^{-\eta_X})^{-1} - \delta \bar{X} \label{eq:X_scaled},
\end{align}
where we have suppressed the dependence of $\bar{\beta}_X$ on $\bar{C}$ in the second equation.
We note that these equations yield biologically meaningful solutions when $\bar{C}>0$ and $\bar{X}>0$, since these variables represent concentrations of proteins in the cell.
In addition, the rate parameters $\bar{\alpha}$,  $\bar{\beta}_C$, $\bar{\beta}_X$, and $\delta$ as well as the concentrations $K_C$ and $K_X$ are all larger than zero.

We proceed to analyze the fixed points of this system of equations and their stability. Because gene $X$ does not regulate gene $C$, the existence and stability of fixed points for $C$ will not depend on $X$. Furthermore, the results for \cref{eq:crp_scaled} will be immediately applicable to \cref{eq:X_scaled}, since $\bar{C}$ will be time-independent at a fixed point. For these reasons, we suppress the subscript $C$ in the following analysis to simplify the notation, and we emphasize that $\bar{C}$ is interchangeable with $\bar{X}$.

We proceed by setting $f(\bar{C})=0$ to obtain a polynomial equation for the fixed points of $\bar{C}$:
\begin{equation}
    \left(\bar{\alpha} + \bar{\beta}-\delta\bar{C}\right)\bar{C}^{\eta} = \delta \bar{C} - \bar{\alpha}. \label{eq:fp_poly}
\end{equation}
This equation has one or three solutions for biologically meaningful values of the parameters.
By setting $df/d\bar{C}=0$, we obtain the condition
\begin{equation}
    \left(\bar{\alpha} + \bar{\beta}-\delta\bar{C}\right)\eta \bar{C}^{\eta-1} - \delta \bar{C}^{\eta} = \delta. \label{eq:fp_poly_der}
\end{equation}
When this equation 
is satisfied simultaneously with \cref{eq:fp_poly} 
a pitchfork bifurcation occurs at a critical value $\bar{C}^*$.

This critical value may be obtained from the following steps:
\begin{enumerate}
    \item[(i)] adding $\delta \bar{C}^{\eta_C}$ to each side of \cref{eq:fp_poly_der},
    \item[(ii)] dividing the result by \cref{eq:fp_poly} to obtain a self-consistent equation for $1 + \bar{C}^{\eta_C}$ in terms of $\bar{C}$, and
    \item[(iii)] using the self-consistency result to substitute out terms with $\bar{C}^{\eta_C}$ in favor of $\bar{C}$ to obtain a quadratic equation in $\bar{C}$.
\end{enumerate} 
The resulting equation is
\begin{equation}
    \frac{1}{2}(\bar{C}^*)^2 - \frac{2\bar{\alpha}\eta_C + \bar{\beta}(\eta_C-1)}{2\delta \eta_C}\bar{C}^*+\frac{\bar{\alpha}(\bar{\alpha}+\bar{\beta})}{2\delta^2}=0, \label{eq:quadratic_cbar}
\end{equation}
which has roots at
\begin{equation}
    \bar{C}^* = \frac{2\bar{\alpha}\eta + \bar{\beta}(\eta-1)}{2\delta \eta} \left(1 \pm \sqrt{1-\frac{\bar{\alpha}(\bar{\alpha}+\bar{\beta})}{\left(2\bar{\alpha}\eta + \bar{\beta}(\eta-1)\right)^2}} \right). \label{eq:cstar_roots}
\end{equation}

For a given value of $\eta>1$, there is a critical value of $\bar{\alpha}$ and $\bar{\beta}$ where multistability arises. This occurs when $f$, $df/d\bar{C}$, and $d^2 f/ d\bar{C}^2$ all equal zero for the same value of $\bar{C}$. Setting $d^2 f/ d\bar{C}^2 = 0$, we obtain
\begin{align}
    \frac{d^2 f}{d\bar{C}^2}= 0 & = \frac{\bar{\beta} \bar{C}^{\eta - 2} \eta \left(- \bar{C}^{\eta} \eta - \bar{C}^{\eta} + \eta - 1\right)}{\left(1+\bar{C}^{\eta}\right)^3},\quad \mbox{which implies} \nonumber \\
    \bar{C}_{\rm crit} & = \left(\frac{\eta-1}{\eta+1}\right)^{\frac{1}{\eta}} \label{eq:C_crit}.
\end{align}
Notably, $\bar{C}_{\rm crit}$ is independent of $\bar{\beta}$. Substituting this value of $\bar{C}$ into $df/d\bar{C}=0$ yields
\begin{align}
    \frac{d f}{d\bar{C}}= 0 & = -\frac{\bar{\beta} \bar{C}^{2 \eta} \eta}{\bar{C} \left(\bar{C}^{\eta} + 1\right)^{2}} + \frac{\bar{\beta} \bar{C}^{\eta} \eta}{\bar{C} \left(\bar{C}^{\eta} + 1\right)} - \delta
    \bar{\beta}, \quad \mbox{which reduces to} \nonumber \\ 
    \bar{\beta}_{\rm crit} & = \frac{4\eta \delta}{(\eta-1)^{\frac{\eta-1}{\eta}}(\eta+1)^{\frac{\eta+1}{\eta}}}, \label{eq:b_crit}
\end{align}
after substituting in \cref{eq:C_crit} for $\bar{C}$.
Substituting \cref{eq:C_crit,eq:b_crit} into \cref{eq:fp_poly} results in  
\begin{equation}
    \bar{\alpha}_{\rm crit} =\frac{\left(\eta-1\right)^2}{4\eta} \bar{\beta}.   \label{eq:a_crit} 
\end{equation}
With these critical values determined, we can determine the region of ($\bar{\alpha}$,$\bar{\beta}$,$\eta$)-space that allows for multiple stable fixed points. 

\begin{figure}[h!]\small\centering
\includegraphics[width=\textwidth]{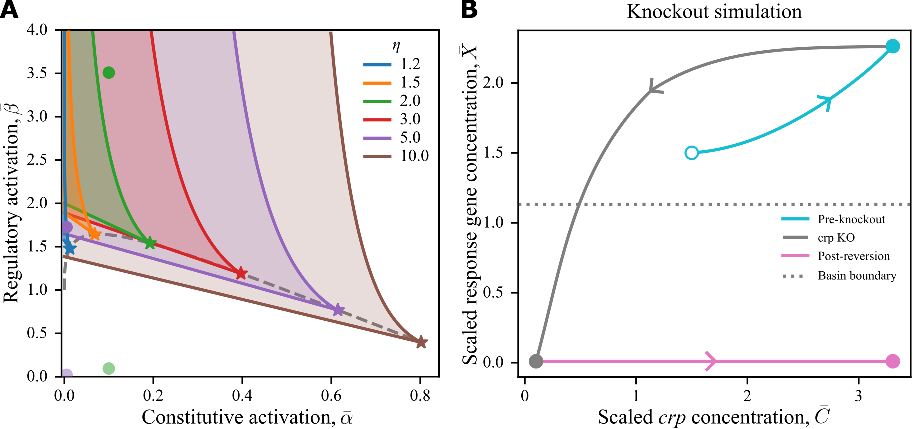}
\caption{  \textbf{Irreversibility in differential equation models of a the key motif in \cref{fig:wt_attr_analysis}.} 
(\textbf{A}) Multistability regions in the constitutive activation--regulatory activation plane, color-coded by the values the Hill coefficient ($\eta$) indicated in the legend. Stars indicate the critical point (derived from \cref{eq:b_crit,eq:a_crit}) for each $\eta$, while the gray dashed line shows the trajectory of the critical point as $\eta$ is continuously increased from $1$ to $10$. Circles indicate exemplar parameters for the constitutive and regulatory activation of \textit{crp} (green, indicating $\eta_C=2$) and the response gene (purple, indicating $\eta_X=5$) when \textit{crp} is active (bold) and knocked out (faded). 
(\textbf{B}) Concentrations of \textit{crp} and the response gene prior to (teal), during (gray), and after \textit{crp} KO (pink), where arrows indicate the direction of time. During each phase, the system is integrated until the system reaches the stable state. The gray dotted line indicates the basin boundary (unstable fixed point in the response-gene concentration) when \textit{crp} is active.
}
\label{fig:diffeq_simulations}
\end{figure}

\Cref{fig:diffeq_simulations}A illustrates the regions of parameter space for which the system is multistable. These regions are determined by fixing $\eta$, calculating the critical point, then progressively increasing $\bar{\beta}$.
For each value of $\bar{\beta}$, we adjust $\bar{\alpha}$ from $\bar{\alpha}_{\rm crit}$ to $\bar{\alpha}_{\rm crit} - (\bar{\beta}-\bar{\beta}_{\rm crit}) \left(1+\bar{C}^{-\eta}_{\rm crit}\right)^{-1}$ in $f(\bar{C})$. 
Next, we solve for the concentrations $\bar{C}^{+}$ and $\bar{C}^{-}$ that maximize and minimize $f(\bar{C})$ by finding the roots of $f'(\bar{C})$. 
Then, we compute $f(\bar{C}^{+})$ and $f(\bar{C}^{-})$ with the adjusted value of $\alpha$. 
This procedure yields the largest and smallest values of $\bar{\alpha}$ for this value of $\bar{\beta}$, which are
\begin{align}
    \bar{\alpha}_{\rm max} (\bar{\beta}) = & \bar{\alpha}_{\rm crit} - (\bar{\beta}-\bar{\beta}_{\rm crit}) \left(1+\bar{C}^{-\eta}_{\rm crit}\right)^{-1} - f(\bar{C}^{-}),\ \mbox{and}\label{eq:a_max_beta}\\
    \bar{\alpha}_{\rm min} (\bar{\beta}) = & \bar{\alpha}_{\rm crit} - (\bar{\beta}-\bar{\beta}_{\rm crit}) \left(1+\bar{C}^{-\eta}_{\rm crit}\right)^{-1} - f(\bar{C}^{+}). \label{eq:a_min_beta}
\end{align}
Since only $\bar{\alpha}\geq 0$ are physically meaningful, we constrain  $\bar{\alpha}_{\rm min} \geq 0$.
We also plot the critical points (determined from output \cref{eq:a_crit} and \cref{eq:b_crit}) parametrically as a function of $\eta \in [1,10]$. From this analysis, it is possible to identify parameters suitable for realizing irreversibility as envisioned by the Boolean model, as shown next.

\paragraph*{Simulations:} 
\Cref{fig:diffeq_simulations}B shows an example of a transient knockout of \textit{crp} causing the irreversible response of gene $\bar{X}$ from simulation of \cref{eq:crp_scaled,eq:X_scaled}.
The initial parameters for \textit{crp} and the response gene are indicated in \cref{fig:diffeq_simulations}A by the bold green and purple circles, respectively (the colors correspond to the value of $\eta$ in the legend). By \cref{eq:dim}, the regulatory activation for $\bar{X}$ already includes the impact of \textit{crp} regulation.
This choice of parameters has a single fixed point for $\bar{C}$, which is highly expressed. The high expression of $\bar{C}$ coupled with the high initial expression of $\bar{X}$ cause the system to evolve toward a state with high concentrations of both genes. When \textit{crp} is subsequently knocked out (implemented by decreasing $\bar{\beta}$ to 0.1 as indicated by the faded green circle in \cref{fig:diffeq_simulations}A), the gene concentrations evolve toward a fixed point with both concentrations near their (low) $\bar{\alpha}$. 
Upon restoration of the regulatory activation of \textit{crp}, its concentration increases. 
The concentration $\bar{X}$, however, remains low because there is a fixed point for \cref{eq:X_scaled} with a low $\bar{X}$.
Although this is only one choice of parameters, it is clear that there are many possible choices that yield concentration trajectories similar to those observed in \cref{fig:diffeq_simulations}B.

\subsubsection*{Examples of potential experiments}

The results of the differential equation modeling help refine the selection of candidate irreversible perturbations and irreversible response genes by allowing us to evaluate their probability of exhibiting irreversibility using qualitative knowledge about the activation strength.
From the transcriptional data we collected, it appears that \textit{crp} has a fairly large constitutive activation since it remains expressed at detectable levels even when \textit{E. coli} is cultivated with glucose as the primary carbon source (which is known to inhibit \textit{crp}). A larger constitutive activation restricts the values of regulatory activation (and Hill coefficient) that allow \textit{crp} to exhibit multistability (\cref{fig:diffeq_simulations}A). 
Since we want \textit{crp} expression to be restored after the KO is removed, \textit{monostability} of \cref{eq:crp_scaled} is desirable so that $\bar{C}$ will spontaneously increase once $\bar{\beta}$ is increased. 
Concerning the response gene $X$, irreversibility requires it to have parameters that admit multistability when \textit{crp} is active. This is facilitated by smaller constitutive activations, larger regulatory activations, and larger Hill coefficients. 
In most cases, larger values of the regulatory activation and Hill coefficient for \textit{crp} also facilitate multistability. 
Under these conditions, the drop in expression in \textit{crp} due to the KO causes a drop in expression of the response gene and restoration of \textit{crp} expression fails to restore expression of the response gene as explained in \cref{fig:diffeq_simulations}B.

From these considerations, one can imagine an experiment in which initially high \textit{crp} expression is achieved by cultivating \textit{E. coli} in glycerol. 
The regulated gene may also be manipulated to be highly expressed, for example by including zinc in the cultivation conditions to ensure that expression of \textit{zraR} is initially large as well.
Then, the reduction in regulatory activation of \textit{crp} can be achieved by inducing CRISPR-interference. 
Subsequently un-inducing CRISPR should allow the \textit{crp} expression to recover. 
However, if the concentration of zinc is low enough, the regulatory activation of \textit{zraR} can be tuned to be in the multistable region.
Because \textit{zraR} expression should be low during the induction of CRISPR, it should not be able to recover even though \textit{crp} expression does.

One key assumption in \cref{eq:crp_unscaled,eq:X_unscaled} is that the regulatory logic is qualitatively similar to a Boolean AND operator. 
Although this information is difficult to ascertain from the literature, it may be possible to infer cases in which the regulation is more likely to be AND based on the location of the transcription factor binding sites to DNA.
If these sites are overlapping, then it is less likely that the regulatory logic is AND, whereas if \textit{CRP} and the regulated protein form a dimer, the logic is more likely to be AND. 
If the regulatory rule is qualitatively an OR function, then \cref{eq:X_unscaled} must be changed accordingly. 
Irreversibility may still be possible, but this would require employing an OE perturbation.

As shown in \cref{fig:diffeq_simulations}A, genes with stronger self-activation and larger Hill coefficients (i.e., more switch-like behavior) tend to have wider regions of multistability. 
Thus, the literature can be consulted to identify irreversible response gene candidates that possess these attributes.
We also note that several genes may have activation parameters that can be altered via the inclusion of specific metabolites (e.g., zinc for the case of \textit{zraR} above). 
The inclusion of these chemicals may be tuned to obtain a regulatory activation strength that admits multistability and to ensure that the response gene is initially ON.
An analogous strategy has been previously employed to show multistability in inducible sugar utilization~\cite{Afroz2014}.

Finally, the response may be heterogeneous among cells. Such heterogeneity could occur, for example, as a result of the stochastic allocation of cellular components during cytokinesis.
Since single-cell sequencing techniques in bacteria are still in their infancy, other approaches, like super-resolution fluorescent microscopy techniques, would be required to observe a heterogeneous response.
The considerations in this section successfully narrow the scope of candidate irreversible perturbations and their associated response genes, but still leave substantial work to develop the required molecular biology tools implement the transient perturbation, ascertain the nature of the regulation, and detect the expression in time.

\clearpage
\newpage
\subsection*{Supplementary Figures}
\phantom{.}
\begin{figure}[h!]\small\centering
\vspace{-1cm}
\includegraphics[width=0.67\textwidth]{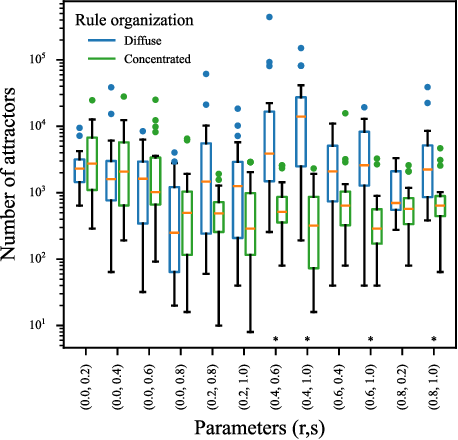}
\caption{ \textbf{Number of attractors across realizations.} Boxplots of the number of attractors across the $M=20$ realizations compare the number under the diffuse (blue) and concentrated (green) rule organization. Asterisks indicate parameters $(r, s)$ for which the diffuse organization has a signficantly larger number of attractors as quantified by the Kruskal-Wallis test with a Bonferroni-corrected $p$-value $< 0.004$. The boxes, orange lines, upper whiskers, and lower whiskers denote the interquartile range, median, maximum (excluding fliers), and minimum,  respectively. Fliers, which are data points that are more than 1.5 times the interquartile range above the median, are plotted separately.
} 
\label{fig:num_attractors}
\end{figure}
\newpage

\begin{figure}[h!]\small\centering
\vspace{-1cm}
\includegraphics[width=0.53\textheight]{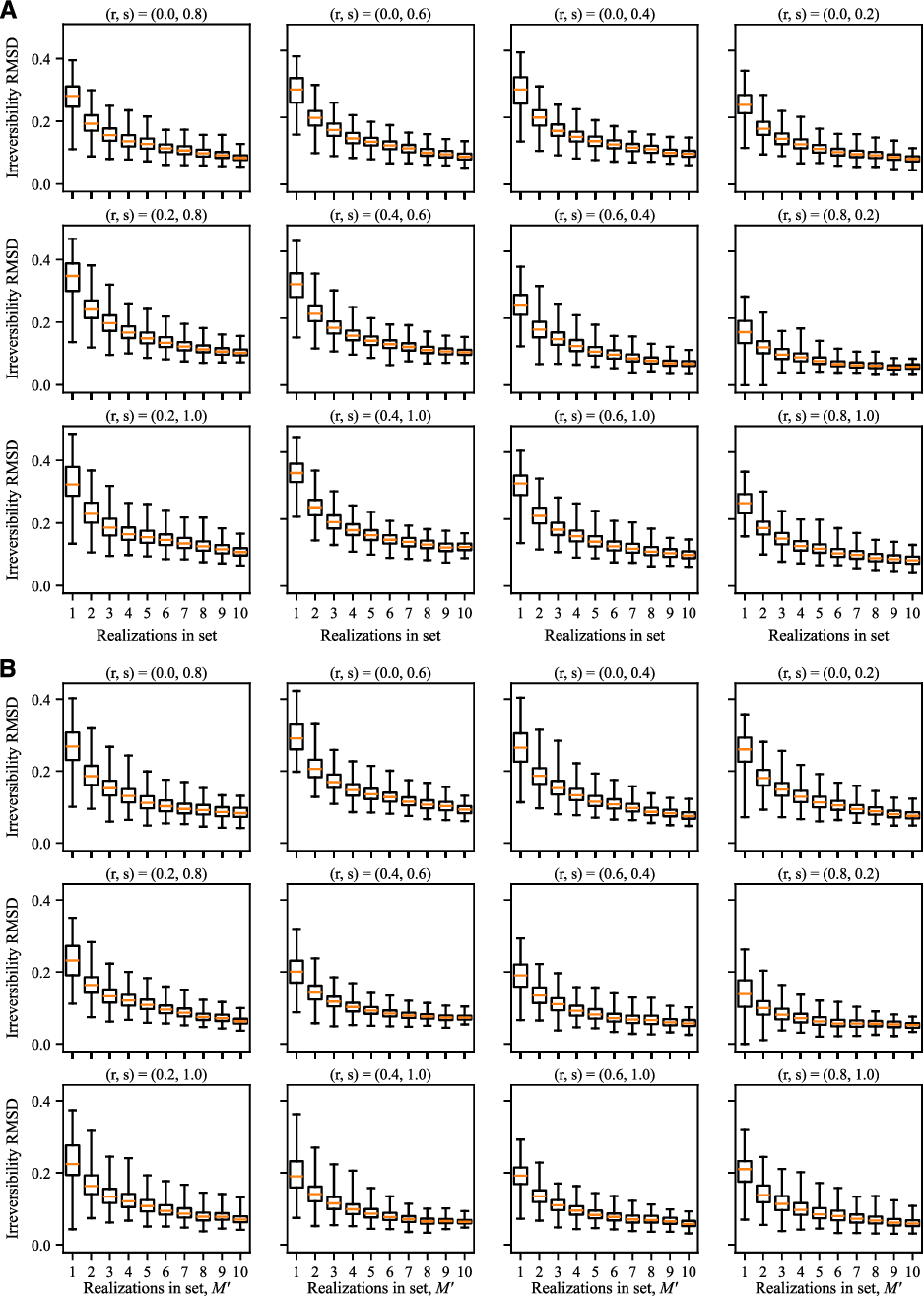}
\caption{ \textbf{Convergence of irreversibility estimated from independent sets of realizations.} (\textbf{A})  Boxplots of the RMSD of the irreversibility probability across $\min(Z(M,M'), 1{,}000)$ independent pairs of realizations of the rules for the ascending input sorting  (diffuse control). (\textbf{B}) Same quantities as in (A) for the descending input sorting (concentrated control). The values of the bias ($s$) and nestedness ($r$) parameters are indicated above each panel. 
 Boxplot symbols retain their meanings from \cref{fig:num_attractors} (there are no fliers).}
\label{fig:rules_ensemble}
\end{figure}
\newpage
\begin{figure}[h!]\small\centering
\includegraphics[width=\textwidth]{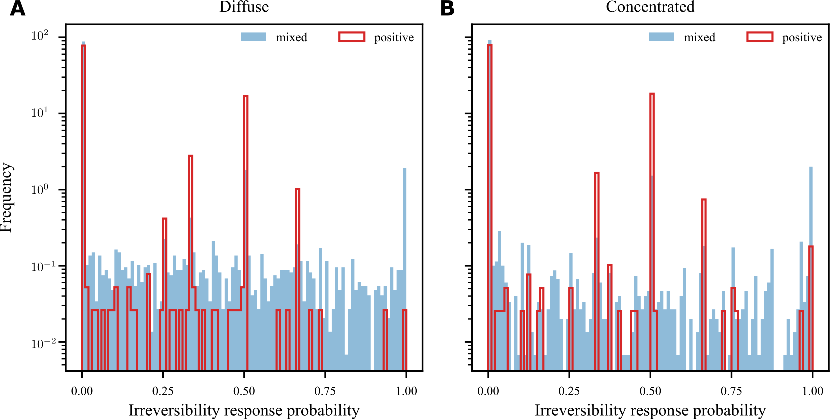}
\caption{ \textbf{Comparison of irreversible response probability depending composition of circuits in the SCC containing the irreversible response gene.} (\textbf{A}) Histograms of the irreversibility response probability across all network ensemble parameters $(r, s)$ and perturbations of \textit{crp} under the diffuse control scenario. Genes are divided into groups by whether they belong to an SCC with only positive circuits (red) or a mixture of positive and negative circuits (blue).  (\textbf{B}) Same as (A), but for the concentrated control scenario.
}
\label{fig:mixed_v_pos}
\end{figure}
\newpage
\begin{figure}
    \centering
    \includegraphics{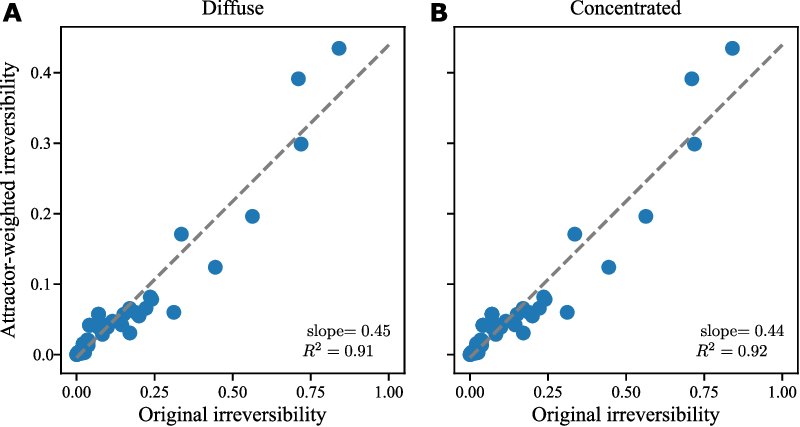}
    \caption{ \textbf{Comparison of irreversibility when weighting the results by attractor basin size versus weighting them uniformly by attractor.} (\textbf{A}) Scatter plot of the irreversibility of each gene averaged over perturbation types, rules realizations, and attractors in the diffuse control scenario (ascending input ordering). The gray dashed line is a linear fit to the data, whose slope indicates the quantitative change in irreversibility due to attractor weighting. The $R^2$ indicates the value of the coefficient of determination. (\textbf{B}) Same as (A), but for the  concentrated control scenario.}
    \label{fig:basinwt_analysis}
\end{figure}
\clearpage
\newpage
\newpage
\newpage

\clearpage
\newpage
\subsection*{Supplementary Tables}
\begin{table}[h]
\footnotesize
    \centering
    \caption{\textbf{Statistics of the  origons with core size $\geq 30$.}}
    \begin{tabular}{>{\raggedright}p{2cm} >{\raggedleft}p{2cm} >{\raggedleft}p{2cm} >{\raggedleft\arraybackslash}p{2.5cm} }
    \hline
        Origon root & \makecell[r]{Total size} & \makecell[r]{Core size}  &  \makecell[r]{Overlap with \\ \emph{phoB} core}     \\ 
         \hline
       \emph{phoB}  & 1406 &	87 &	87 \\
        \emph{nsrR} & 1399 & 88 & 81 \\
        \emph{acrR} & 941 & 59 & 58 \\
        \emph{cpxR} & 550 & 35 & 33 \\
        \emph{phoP} & 535 & 32 & 29 \\
        \emph{slyA} & 476 & 30 & 29 \\
        \emph{rutR} & 475 & 31 & 29 \\
        \emph{basR} & 474 & 32 & 30 \\
        \emph{torR} & 468 & 30 & 29 \\
        \emph{rcdA} & 466 & 30 & 29 \\
        \emph{sdiA} & 464 & 30 & 29 \\
        \emph{sutR} & 463 & 30 & 29 \\
        \emph{ecpR} & 463 & 30 & 29 \\
        \emph{lrhA} & 462 & 30 & 29 \\
        \hline
    \end{tabular}
    \label{tab:origon_overlap}
\end{table}

\newpage
\clearpage

\begin{table}[h]
\footnotesize
    \centering
    \caption{\textbf{Irreversibility of genes with log-fold change $>0.5$ when batch cultivated following adaptive evolution to \textit{crp} KO.}}
    \begin{tabular}{l r r c r c r}
\hline Gene&Irreversible response probability&Shortest path from \textit{crp}&$\sigma_u^{\rm mod}$&$k_u^+$&Autoregulation&$\ln \rho_u / \langle \rho \rangle$  \\
\hline\textit{melR}&0.22&1&$+$&2&$+$&$-$3.11 \\
\textit{flhD}&0.02&1&$+$&9&$0$&$-$1.72 \\
\textit{malI}&0.00&1&$+$&2&$-$&$-$1.65 \\
\textit{rcsA}&0.03&2&$-$&4&$+$&1.42 \\
\textit{rhaR}&0.19&1&$+$&3&$+$&$-$1.24 \\
\textit{flhC}&0.01&1&$+$&9&$0$&$-$1.17 \\
\textit{tdcA}&0.04&1&$+$&3&$+$&$-$1.14 \\
\textit{adiY}&0.03&3&$\pm$&1&$0$&$-$0.66 \\
\textit{bglJ}&0.03&3&$\pm$&2&$0$&$-$0.64 \\
\textit{ptsG}&0.02&1&$+$&6&$-$&0.62 \\
\textit{lsrR}&0.00&1&$+$&2&$-$&$-$0.61  \\
 \hline  \end{tabular}
    \label{tab:batch_adapt}
\end{table}

\clearpage
\newpage

\begin{table}[t]
\footnotesize
    \centering
    \caption{\textbf{Irreversibility of genes with log-fold change $>0.5$ when chemostat cultivated following adaptive evolution to \textit{crp} KO.}}
    \begin{tabular}{l r r c r c r}
\hline Gene&Irreversible response probability&Shortest path from \textit{crp}&$\sigma_u^{\rm mod}$&$k_u^+$&Autoregulation&$\ln \rho_u / \langle \rho \rangle$  \\
\hline \textit{tdcA}&0.04&1&$+$&3&$+$&$-$3.65\\
\textit{melR}&0.22&1&$+$&2&$+$&$-$3.02\\
\textit{flhD}&0.02&1&$+$&9&$0$&$-$2.48\\
\textit{aidB}&0.01&3&$+$&2&$-$&2.46\\
\textit{rhaR}&0.19&1&$+$&3&$+$&$-$2.38\\
\textit{flhC}&0.01&1&$+$&9&$0$&$-$2.16\\
\textit{glcC}&0.00&1&$+$&4&$-$&$-$2.12\\
\textit{gadE}&0.02&1&$-$&10&$+$&1.98\\
\textit{lsrR}&0.00&1&$+$&2&$-$&$-$1.71\\
\textit{ydeO}&0.03&2&$-$&7&$-$&$-$1.66\\
\textit{fucR}&0.20&1&$+$&2&$+$&$-$1.64\\
\textit{prpR}&0.00&1&$+$&3&$-$&$-$1.64\\
\textit{galS}&0.04&1&$+$&3&$-$&$-$1.59\\
\textit{hns}&0.03&2&$\pm$&4&$-$&$-$1.51\\
\textit{leuO}&0.02&3&$\pm$&5&$+$&$-$1.47\\
\textit{pdeL}&0.00&NA&NA&2&$+$&$-$1.40\\
\textit{yeiL}&0.02&3&$\pm$&3&$+$&$-$1.31\\
\textit{fur}&$<$0.01&1&$+$&4&$-$&$-$1.23\\
\textit{mlc}&0.01&1&$-$&3&$-$&$-$1.15\\
\textit{malI}&0.00&1&$+$&2&$-$&$-$1.12\\
\textit{yjjQ}&0.03&3&$\pm$&2&$0$&$-$1.09\\
\textit{gadW}&0.02&2&$\pm$&8&$-$&1.08\\
\textit{bglJ}&0.03&3&$\pm$&2&$0$&$-$1.07\\
\textit{evgA}&0.03&3&$\pm$&2&$+$&$-$0.98\\
\textit{srlR}&0.01&1&$+$&4&$-$&$-$0.96\\
\textit{rbsR}&0.00&1&$+$&2&$-$&$-$0.91\\
\textit{cspA}&0.03&2&$+$&2&$0$&$-$0.91\\
\textit{dcuR}&0.04&1&$+$&3&$0$&$-$0.80\\
\textit{metR}&$<$0.01&3&$+$&2&$-$&$-$0.79\\
\textit{fnr}&0.10&2&$-$&3&$-$&$-$0.78\\
\textit{narL}&0.09&3&$+$&1&$0$&0.75\\
\textit{gadX}&0.03&1&$-$&13&$+$&0.72\\
\textit{lldR}&0.05&4&$+$&2&$-$&$-$0.69\\
\textit{rhaS}&0.19&1&$+$&3&$+$&$-$0.67\\
\textit{nhaR}&0.03&3&$\pm$&2&$+$&0.64\\
\textit{fliZ}&0.02&2&$\pm$&4&$0$&$-$0.59\\
\textit{purR}&$<$0.01&2&$-$&2&$-$&$-$0.56\\
\textit{yqjI}&0.04&3&$-$&2&$-$&0.55\\
\textit{cra}&0.00&NA&NA&1&$0$&$-$0.55\\
\textit{glnG}&0.00&1&$+$&3&$-$&$-$0.55\\
\textit{nac}&0.03&2&$\pm$&4&$-$&$-$0.54\\
\textit{fhlA}&0.01&3&$-$&2&$+$&0.51 \\ \hline  \end{tabular}
    \label{tab:chemo_adapt}
\end{table}

\end{document}